\documentclass[twocolumn]{aastex701}

\usepackage{amsmath}
\usepackage{xcolor}
\usepackage{caption}
\submitjournal{PASP}
\graphicspath{{./}}

\begin{document}

\title{Honoring Paris Pi\c{s}mi\c{s}’ Legacy: Uniform Astrophysical Properties of 14 Understudied Clusters from Gaia DR3}

\author[orcid=0000-0001-7420-0994,sname='Koç']{Seliz Koç}
\affiliation{Department of Data Science and Analytics, Faculty of Engineering and Natural Sciences, Istanbul Galata University, 34440 Beyo\u{g}lu, Istanbul, T\"{u}rkiye}
\email{seliz.koc@galata.edu.tr}

\author[orcid=0000-0001-9809-7493,sname='Alan']{Neslihan Alan}
\affiliation{Fatih Sultan Mehmet Vakif University, Department of History of Science, 34664, Istanbul, T\"{u}rkiye}
\email{nalan@fsm.edu.tr}

\author[orcid=0000-0002-9993-7244,sname='Akbaba']{Furkan Akbaba}
\affiliation{Institute of Graduate Studies in Science, Istanbul University, Istanbul, Turkey}
\email{furkan.akbaba@ogr.iu.edu.tr}

\author[orcid=0000-0002-0435-4493,sname='Plevne']{Olcay Plevne}
\affiliation{Faculty of Science, Department of Astronomy and Space Sciences,Istanbul University, Istanbul, Turkey}
\email[show]{olcayplevne@istanbul.edu.tr}

\begin{abstract}
Open clusters are fundamental tracers of the chemical and dynamical evolution of the Galactic disc. This study presents a systematic characterisation of 14 understudied open clusters from the Pismis catalogue, selected as those for which the \textit{Gaia} DR3 data support a well-defined stellar sequence, and spanning a wide range of ages and distances. Utilizing high-precision Gaia DR3 data, we aim to provide a high-fidelity modernisation of these historical discoveries while testing the consistency of our analytical framework across diverse stellar populations. Membership probabilities were determined via the unsupervised UPMASK algorithm using \textit{Gaia} DR3 astrometric data (positions, proper motions, and parallax). To ensure high sample purity, the member samples entering the isochrone fits were restricted to $P \geq 0.90$, with all subsequent astrophysical parameters calculated through membership-weighted likelihoods. Fundamental properties—including age, distance, reddening, and metallicity—were derived using a Bayesian Nested Sampling approach to isochrone fitting, a methodology chosen for its robustness in navigating the parameter degeneracies typical of dense Galactic fields. 
The 14 systems exhibit a broad distribution of properties, with ages ranging approximately 11 Myr to 1.5 Gyr and distances spanning approximately 690 to 5445 pc. These refined and uniform parameters reconcile historical observations with the precision era of modern astrophysics, providing an internally homogeneous baseline for systems that large-scale automated surveys recover only partially. A direct cross-match against widely used automated catalogues shows that they typically contain only about a third of the members identified here, and substantially less in the most crowded fields, underscoring the value of cluster-specific analysis. This work contributes to a more complete census of the Galactic disc and sets out a reproducible, per-cluster reliability assessment that can be applied to the many comparable low-latitude systems that remain to be characterised.
\end{abstract}


\section{Introduction}

Open star clusters are widely used to study the structure and evolution of the Milky Way disk \citep[e.g.,][]{Friel1995, Dias_2002, Cantat-Gaudin_2020}. Because the member stars within a single system form simultaneously from the collapse of the same giant molecular cloud, they share a common age, distance, and initial chemical composition \citep{Lada_2003, Portegies-Zwart10, Kim_2017}. This physical homogeneity reduces the number of free parameters, making open clusters as a whole --- not merely the main-sequence morphology of an individual system --- reliable benchmarks for testing stellar evolution models. Located close to the Galactic plane, their spatial distribution provides a direct way to trace the large-scale layout of the Galactic disk \citep{Friel1995, Cantat-Gaudin_2020, Castro-Ginard2022}. They are, in effect, the only population that simultaneously carries a reliable age, a reliable distance, and a reliable composition, which is why they remain the primary calibrators of the disc age--metallicity relation, of the radial abundance gradient and its evolution, and of cluster dissolution timescales \citep{Krumholz_2019, Netopil_2022, Spina22, Akbaba2024, Horta2026}.

The systematic study of these systems is old enough to have passed through several distinct observational regimes. The visual era, culminating in the survey of the southern sky by \citet{Herschel1847}, established which condensations existed but said little about what they were. Physical characterisation began with photographic photometry: \citet{Trumpler_1930} used it to place clusters on a common distance scale, to introduce the concentration classes still in use, and --- through the systematic reddening of distant clusters --- to establish interstellar absorption itself. The following decades were dominated by wide-field photographic surveys of the southern Milky Way, which is where most previously uncatalogued clusters were still hiding; the classification schemes and compilations that organised these discoveries \citep[e.g.,][]{Ruprecht_1966} defined the working census of open clusters for the rest of the century.

The Pismis catalogue has an established place among these surveys. It is named after Marie Paris Pi\c{s}mi\c{s} (1911--1999)\footnote{One of the earliest prominent female astronomers globally, Pi\c{s}mi\c{s} was the first woman to earn a Ph.D. in mathematics from Istanbul University, T\"urkiye. She pioneered astronomy research in T\"urkiye and subsequently made pivotal contributions to shaping the infrastructure of modern Mexican astrophysics.}, who, analysing wide-field photographic plates obtained with the Schmidt camera of the Tonantzintla Observatory, Mexico, identified 24 open clusters and two globular clusters, introducing the Pismis catalogue to the literature \citep{Pismis1959}. The search was deliberately directed at low Galactic latitudes, towards the inner disc and the third quadrant, precisely where crowding and extinction had kept earlier surveys from being complete. That choice is the reason the catalogue exists, and, as shown below, it is also the reason its members have remained difficult objects ever since.

The observational regime changed again with the arrival of all-sky astrometry. The \textit{Gaia} space mission \citep{Gaia16}, particularly Data Release 3 \citep[DR3;][]{Gaia_DR3}, significantly increased the astrometric and photometric precision available for stellar cluster analysis. Parallaxes and proper motions of individual stars turned membership determination from a photometric plausibility argument into a measurement, and unsupervised clustering applied to the resulting astrometric space has both revised the membership of classical clusters and expanded the census by thousands of new candidates \citep{Cantat-Gaudin_2018, Castro-Ginard2022, Hunt_2023, Hunt_2024}. Homogeneous parameter catalogues built on these membership lists \citep[e.g.,][]{Bossini_2019, Cantat-Gaudin_2020, Dias_2021} now cover the disc well enough to trace spiral structure and the warp \citep{Joshi_2016, Poggio_2021}, while the large spectroscopic surveys --- Gaia-ESO, OCCAM/APOGEE, and their successors --- have added chemical abundances for hundreds of clusters and, with them, a far sharper picture of the radial metallicity gradient \citep{Donor_2020, Netopil_2022, Randich_2022, Spina22}. The scale of the census has in turn made automated parameter estimation a necessity, and pipelines that fit thousands of clusters in a single pass --- whether by neural networks trained on synthetic populations \citep{Cavallo_2024} or by Bayesian nested sampling applied cluster by cluster \citep{Akbaba2026, Plevne2026} --- now supply ages, distances and extinctions for the bulk of the known population. Taken together, these programmes have made the open-cluster population a statistically usable tracer of the disc for the first time.

The strength of these surveys, however, is statistical, and it is bought at a price that is paid unevenly across the sky. While global automated surveys catalog thousands of open clusters, their effectiveness decreases for heavily reddened targets, sparse systems, or dense fields toward the inner Galactic disk, a limitation we quantify directly for the present sample in Section~\ref{subsec:comparison}. Different automated pipelines \citep[e.g.,][]{Dias_2002, Kharchenko_2013, Sampedro_2017, Cantat-Gaudin_2020, Hunt_2023} therefore often yield discrepant fundamental parameters for the same clusters. These variations arise from different membership selection criteria, the handling of faint stars, and the inherent age--distance--reddening degeneracy, the latter of which we discuss concretely for a subset of the present sample in Section~\ref{sec:fit_reliability}. A uniform pipeline must adopt uniform thresholds --- on field density, on the number of recovered members, on the tolerated contamination --- and objects that sit near those thresholds are either assigned truncated membership lists or dropped altogether. Because the thresholds are set by what works for the bulk of the population, the systems that fail them are not a random subset: they are concentrated at low $|b|$, at high extinction, and towards the inner disc. Cluster-specific analysis is required to resolve these discrepancies, and it is exactly in this regime that it is still required.

The Pismis clusters fall squarely in that regime, and this is what makes them worth revisiting now. Despite their historical importance, a significant portion of the Pismis open clusters has lacked homogeneous investigation. While specific objects have been analyzed through dedicated multi-wavelength studies \citep[e.g.,][]{Carraro2012, Hatzidimitriou_2019, Baravalle2021}, many systems still lack a consistent characterization based on high-precision data. They are understudied not by accident but by circumstance: all 14 of the targets analysed here lie within $|b| \leq 3\fdg21$ of the Galactic plane and thirteen suffer moderate-to-high extinction ($A_V > 1$~mag, reaching $3.7$~mag for Pismis~3), placing the sample in precisely the crowded, heavily-reddened inner and third-quadrant disc where automated catalogues are least reliable. This coincidence turns a tribute into a controlled experiment. Rather than adding one more parameter catalogue --- a task already addressed by wide-field surveys --- we use these difficult systems to ask a specific, quantitative question: in the crowded, reddened disc, how much does cluster-specific membership analysis change the recovered stellar population relative to automated pipelines, how does that difference propagate into the inferred astrophysical parameters, and which of those parameters can be trusted?

To honor the memory and scientific legacy of Paris Pi\c{s}mi\c{s}, this study provides a uniform re-evaluation of a sample of 14 Pismis open clusters selected as suitable for analysis (the selection criteria and the full list of targets are given in Section~\ref{sec:sampleselection}). Our analysis relies on a Bayesian Nested Sampling isochrone fitting approach to derive the fundamental astrophysical parameters of these selected clusters. To constrain this Bayesian framework, we first establish cluster membership from the \textit{Gaia} DR3 database using the unsupervised \textsc{UPMASK} algorithm \citep{Krone-Martins_2014}. This methodology is widely used both in wide-field surveys and in dedicated single-cluster analyses to isolate cluster members from field star contamination \citep[e.g.,][]{Cantat-Gaudin_2020, Castro-Ginard2020, Koc_2023, Nasser_2025, Yontan2026}, including recent studies that combine it with ground-based photometry to characterise individual systems in detail \citep{Cakmak_2024, Karagoz_2025, Tanik2025}. Processing the entire sample through this single pipeline yields an internally homogeneous set of parameters, free of the study-to-study methodological scatter that affects heterogeneous literature compilations, and thus a uniform baseline for comparing these systems within the Galactic disc; we quantify in Section~\ref{subsec:comparison} how far the resulting membership samples and parameters depart from those of the automated catalogues.

We then confront these membership-weighted results, cluster by cluster, both with the major automated catalogues (Section~\ref{subsec:comparison}) and with a quantitative fit-reliability framework of our own (Section~\ref{sec:fit_reliability}). The intended contribution is therefore twofold. First, homogeneous and reproducible fundamental parameters for 14 systems that the modern census has largely passed over, closing a gap left open since the plates were taken. Second, and more generally, a direct measurement of how much automated membership assignment costs in the reddened inner disc, together with an explicit, per-cluster statement of which derived parameters survive that regime and which do not --- a diagnostic that is transferable to the many other low-latitude clusters awaiting the same treatment.

The remainder of this paper is structured as follows. Section~\ref{sec:data} describes the \textit{Gaia} data retrieval and quality filters (Section~\ref{sec:query}), the literature context for the selected clusters (Section~\ref{sec:litcontext}), and the reduction of the original catalogue from 24 to the 14 analysable systems (Section~\ref{sec:sampleselection}). Section~\ref{sec:methods} presents the method: membership determination with \textsc{UPMASK} and the resulting astrometric parameters (Section~\ref{sec:membership}), followed by the Bayesian Nested Sampling isochrone fitting used to derive ages, distances, reddenings, and metallicities (Section~\ref{sec:method}). Section~\ref{sec:results} reports the derived parameters, using Pismis~18 as a worked example of the posterior diagnostics, and compares both our membership samples and our parameters against the major automated catalogues (Section~\ref{subsec:comparison}). Section~\ref{sec:discussion} discusses the results: cluster-by-cluster notes against prior studies, the fit-reliability framework that identifies which parameters can be trusted (Section~\ref{sec:fit_reliability}), the sample in the context of the Galactic disc, and the current state of its spectroscopic coverage (Section~\ref{sec:spec_quality}). The main conclusions are summarised in Section~\ref{sec:conclusion}. Supporting posterior and CMD diagnostics for the full sample are provided as online supplementary material, and the complete parameter table is collected in Appendix~\ref{app:fundamental}.

\section{Data}
\label{sec:data}

\subsection{\textit{Gaia} Data Retrieval and Quality Filters}
\label{sec:query}

Data for the Pismis open clusters were retrieved from the \textit{Gaia} Archive using Data Release 3 \citep[DR3;][]{Gaia_DR3}. For each of the 24 clusters in the original catalogue of \citet{Pismis1959}, we performed a spatial cone search centred on the literature coordinates listed in Table~\ref{tab:pismis_coordinates}, using a radius of 40 arcminutes ($40'$). This search radius sufficiently covers the cluster area while providing an adequate field-star population for the membership assignment described in Section~\ref{sec:membership}. The retrieved dataset comprises five-parameter astrometric solutions---celestial coordinates ($\alpha, \delta$), proper motions ($\mu_{\alpha}\cos\delta, \mu_{\delta}$), and parallax ($\varpi$)---together with $G$, $G_{\rm BP}$, and $G_{\rm RP}$ photometry.

To ensure high astrometric quality, sources were filtered using a Renormalised Unit Weight Error threshold of $\text{RUWE} < 1.4$ \citep{Lindegren_2018}, and only stars with a relative parallax uncertainty $\sigma_{\varpi}/\varpi \leq 0.2$ were retained. Because RUWE increases sharply for astrometric solutions perturbed by an unresolved companion, this cut simultaneously removes a fraction of the unresolved binary population from the sample \citep{Lindegren2021}. While this improves the purity of the single-star sequence used for isochrone fitting, it implies that the resulting membership catalogues are not suited to inferring the intrinsic binary fraction of these clusters, and that our sample is biased against short-period binaries with the most strongly perturbed astrometric solutions.

Photometric analysis and evolutionary modelling were performed using $G$ vs. $(G_{\rm BP}-G_{\rm RP})$ colour--magnitude diagrams (CMDs). To restrict photometric errors to acceptable limits---typically below 0.01~mag in $G$ and 0.1~mag in the colour index---and to mitigate incompleteness at the faint end of the \textit{Gaia} survey, a magnitude limit of $G < 19$~mag was applied across the sample; this restricts the analysis to the magnitude range in which \textit{Gaia} DR3 astrometric and photometric solutions are most reliable in uncrowded fields. Because parallax precision degrades toward fainter magnitudes, this cut acts jointly with the $\sigma_{\varpi}/\varpi \leq 0.2$ criterion to preferentially retain upper-main-sequence and turn-off stars, which dominate the age-sensitive part of the CMD, at the expense of the faintest, most distant candidate members.

In practice, however, the magnitude limit is not what truncates the sample. Applied to the 52\,103 sources that pass the remaining quality criteria across the 14 fields, $41.4\%$ have $G \geq 19$, but $99.8\%$ of those are already removed by the relative parallax criterion $\sigma_{\varpi}/\varpi \leq 0.2$; the magnitude cut alone eliminates 36 sources, or $0.07\%$ of the field sample. The same holds for the members: of the 3464 high-probability members (Section~\ref{sec:membership}) that satisfy every other criterion, only 17 are excluded by $G < 19$ and not already by the parallax criterion, against 805 excluded by the parallax criterion alone. The faint limit of our member catalogues is therefore set by the precision of the \textit{Gaia} astrometry, not by the adopted magnitude cut, and the two are not independent choices: parallax uncertainty grows steeply toward faint magnitudes, so a $20\%$ relative-parallax criterion already implies a magnitude limit of its own. Consistently with this, the faintest retained member lies below $G = 19$ in all 14 clusters, ranging from $G = 17.2$ for Pismis~1 to $G = 19.0$ for Pismis~6.

For the more distant systems 
the limit is additionally imposed by distance itself, independent of the adopted magnitude cut. At the reddening and distance we derive for Pismis~7 ($d = 5367$~pc, $A_G = 2.03$~mag), $G = 19$ corresponds to an absolute magnitude of $M_G \approx 3.3$, that is, to the immediate vicinity of the main-sequence turn-off; the same limit corresponds to $M_G \approx 9.3$ for the nearest cluster, Pismis~4 at $690$~pc. In the distant, reddened clusters the entire observable sequence therefore lies above the magnitude limit, and the low-mass main sequence is inaccessible regardless of where the cut is placed. This is the sense in which our member samples are complete: they contain the age-sensitive part of the CMD in full, and are truncated below it by the depth of the survey rather than by the adopted selection criteria. To maintain physical consistency with the PARSEC stellar evolution models \citep{Bressan_2012} used for isochrone fitting (Section~\ref{sec:method}), we adopted the photometric zero-point corrections and passband definitions of \citet{Riello_2021}. This data preparation enables the reliable determination of fundamental parameters, including age, distance, and reddening, through the Bayesian Nested Sampling approach \citep{Skilling_2004} described in Section~\ref{sec:method}.

\subsection{Literature Context for the Selected Clusters}
\label{sec:litcontext}

The literature record for the 14 selected Pismis clusters is markedly uneven, ranging from systems with essentially no dedicated study to a handful with multiple independent age and distance determinations. Table~\ref{tab:pismis_clusters} compiles the fundamental parameters reported by four representative wide-field \textit{Gaia}-based catalogues \citep{Kharchenko_2013, Cantat-Gaudin_2020, Dias_2021, Hunt_2024}.

Several clusters lack any dedicated individual investigation and are characterised solely through automated survey pipelines: Pismis~4 \citep[Table~\ref{tab:pismis_clusters};][]{Cantat-Gaudin_2020, Dias_2021, Hunt_2024}, and Pismis~14, for which the values reported by \citet{Bukowiecki_2011} and \citet{Kharchenko_2013} show significant discrepancies stemming from the inherent limitations of automated catalogue data \citep{Carraro_2017}. Pismis~1, despite inclusion in global surveys \citep{Cantat-Gaudin_2020, Hunt_2024}, remains poorly constrained owing to its proximity to the Galactic plane and severe line-of-sight contamination \citep{Pismis1959}.

For several other systems, dedicated photometric studies exist but disagree substantially on age, distance, or metallicity. Pismis~3 spans age estimates from $\sim$1.1 to $\sim$3~Gyr and distances from 1.5 to 2.8~kpc across independent analyses \citep{Carraro_1994, Tadross_2008, Bisht2022, Obasi_2025, Zhang_2025}. Pismis~7, 13, and 19 show comparable levels of disagreement between early photometric work \citep{Vogt_1973, Claria_1979, Phelps_1994, Piatti_1998, Ahumada_2005} and more recent \textit{Gaia}-based or CCD studies \citep{Carraro_2004, Giorgi_2005, Cantat-Gaudin_2020, Poggio_2021,Cakmak_2021, Bisht2022}. Pismis~15 and Pismis~18, the two most extensively studied clusters in the sample, have converging age estimates near 0.7--1.4~Gyr \citep{Piatti_1998, Carraro_2005, Tadross_2008, Hatzidimitriou_2019, Pena_Ramirez_2021, Bisht2022, Belwal_2025}, though their metallicity determinations remain in tension (see Section~\ref{sec:individual_clusters}).

Two clusters, Pismis~6 and Pismis~8, have been proposed as a physically associated binary system sharing a common age of order 1--10~Myr \citep{Kopchev_2008}, based on early classifications as young clusters \citep{Vogt_1973, FitzGerald_1979, Forbes_1994, Giorgi_2005}; this putative association remains unconfirmed. Pismis~9, in contrast, has recently been established as an independent foreground cluster unrelated to the nearby NGC~2659 \citep{Giorgi_2023, Hetem_2024}, and Pismis~5 has been linked to a primordial triple-cluster system within the Vela Molecular Cloud \citep{Vogt_1973, Bonatto_2009, Dias_2021, Qin_2023}. Pismis~12, finally, shows unusually consistent literature parameters across near-infrared and \textit{Gaia}-based studies alike \citep{Bica_2011, Bisht2022}.

This heterogeneity in observational coverage and methodology---spanning photographic, 2MASS, and multiple \textit{Gaia} data releases---motivates the uniform, membership-weighted re-analysis presented in this work. A detailed cluster-by-cluster comparison between our results and this literature record is given in Section~\ref{sec:individual_clusters}.

\begin{table*}
\centering
\caption{Literature compilation of fundamental parameters for selected Pismis clusters. Values are listed in chronological order of the references.}
\label{tab:pismis_clusters}
\begin{tabular}{l l l l l}
\hline
Cluster & log(Age) & Distance (pc) & ${[\rm Fe/H]}$ (dex) & $A_V$(mag) \\
\hline

Pismis 1
& 7.800$^{1}$, 7.470$^{2}$, 7.031$^{3}$, 8.180$^{5}$
& 5903$^{1}$, 5666$^{2}$, 4333$^{3}$, 4630$^{5}$
& $-0.194^{3}$
& 1.807$^{1}$, 1.690$^{2}$, 2.044$^{3}$ \\

Pismis 3
& 8.925$^{1}$, 9.500$^{2}$, 8.996$^{3}$, 9.026$^{5}$
& 1418$^{1}$, 2349$^{2}$, 2064$^{3}$, 2112$^{5}$
& $-0.076^{3}$
& 3.680$^{1}$, 2.350$^{2}$, 3.313$^{3}$ \\

Pismis 4
& 8.155$^{1}$, 8.080$^{2}$, 8.094$^{3}$, 8.066$^{5}$
& 575$^{1}$, 695$^{2}$, 691$^{3}$, 681$^{5}$
& $-0.200^{1}$, $-0.089^{3}$
& 0.031$^{1}$, 0.180$^{2}$, 0.292$^{3}$ \\

Pismis 5
& 7.300$^{1}$, 7.000$^{2}$, 6.883$^{3}$, 6.637$^{5}$
& 702$^{1}$, 941$^{2}$, 921$^{3}$, 913$^{5}$
& $-0.034^{3}$
& 2.582$^{1}$, 1.480$^{2}$, 1.388$^{3}$ \\

Pismis 6
& 7.580$^{1}$, 7.410$^{2}$, 7.285$^{3}$, 7.600$^{5}$
& 1668$^{1}$, 1763$^{2}$, 1737$^{3}$, 1720$^{5}$
& $0.126^{3}$
& 1.181$^{1}$, 1.000$^{2}$, 1.401$^{3}$ \\

Pismis 7
& 8.705$^{1}$, 8.890$^{2}$, 8.968$^{3}$, 8.487$^{5}$
& 4783$^{1}$, 4611$^{2}$, 3843$^{3}$, 4947$^{5}$
& $-0.225^{3}$
& 2.905$^{1}$, 1.600$^{2}$, 2.078$^{3}$ \\

Pismis 8
& 7.430$^{1}$, 7.510$^{2}$, 7.137$^{3}$, 7.675$^{5}$
& 1363$^{1}$, 1903$^{2}$, 1756$^{3}$, 1765$^{5}$
& $-0.034^{3}$
& 2.195$^{1}$, 1.580$^{2}$, 2.077$^{3}$ \\

Pismis 9
& 7.360$^{1}$, 8.500$^{2}$, 7.608$^{3}$, 8.458$^{5}$
& 1713$^{1}$, 2120$^{2}$, 1815$^{3}$, 1865$^{5}$
& $-0.080^{3}$
& 1.581$^{1}$, 1.590$^{2}$, 1.528$^{3}$ \\

Pismis 12
& 9.205$^{1}$, 9.090$^{2}$, 9.087$^{3}$, 9.101$^{5}$
& 2221$^{1}$, 2107$^{2}$, 2114$^{3}$, 2094$^{5}$
& $-0.146^{3}$
& 1.615$^{1}$, 1.410$^{2}$, 1.999$^{3}$ \\

Pismis 13
& 7.950$^{1}$, 7.800$^{2}$, 8.285$^{3}$, 7.794$^{5}$
& 2600$^{1}$, 2926$^{2}$, 2367$^{3}$, 2668$^{5}$
& $0.028^{3}$
& 2.046$^{1}$, 1.670$^{2}$, 1.916$^{3}$ \\

Pismis 14
& 8.355$^{1}$, 8.110$^{2}$, 8.185$^{3}$, 8.065$^{5}$
& 1775$^{1}$, 1266$^{2}$, 1234$^{3}$, 1238$^{5}$
& $0.146^{3}$
& 1.485$^{1}$, 0.570$^{2}$, 0.764$^{3}$ \\

Pismis 15
& 9.095$^{1}$, 8.940$^{2}$, 9.258$^{3}$, 8.652$^{5}$
& 2558$^{1}$, 2559$^{2}$, 1872$^{3}$, 2254$^{5}$
& $-0.400^{1}$, $0.456^{3}$, $0.040^{4}$
& 1.550$^{1}$, 1.890$^{2}$, 1.560$^{3}$ \\

Pismis 18
& 8.975$^{1}$, 8.760$^{2}$, 8.832$^{3}$, 8.509$^{5}$
& 2309$^{1}$, 2860$^{2}$, 2189$^{3}$, 2598$^{5}$
& $-0.033^{3}$, $0.14^{4}$
& 1.615$^{1}$, 1.810$^{2}$, 2.165$^{3}$ \\

Pismis 19
& 9.015$^{1}$, 8.920$^{2}$, 8.046$^{3}$, 8.090$^{5}$
& 1978$^{1}$, 3515$^{2}$, 1983$^{3}$, 3034$^{5}$
& $0.064^{3}$
& 4.033$^{1}$, 3.660$^{2}$, 1.743$^{3}$ \\

\hline
\end{tabular}

\vspace{0.3cm}
(1) \cite{Kharchenko_2013};
(2) \cite{Cantat-Gaudin_2020};
(3) \cite{Dias_2021};
(4) \cite{Randich_2022};
(5) \cite{Hunt_2024}.
\end{table*}

\subsection{Sample Selection: From 24 to 14 Clusters}
\label{sec:sampleselection}

\begin{figure*}
    \centering
    \includegraphics[width=0.92\textwidth]{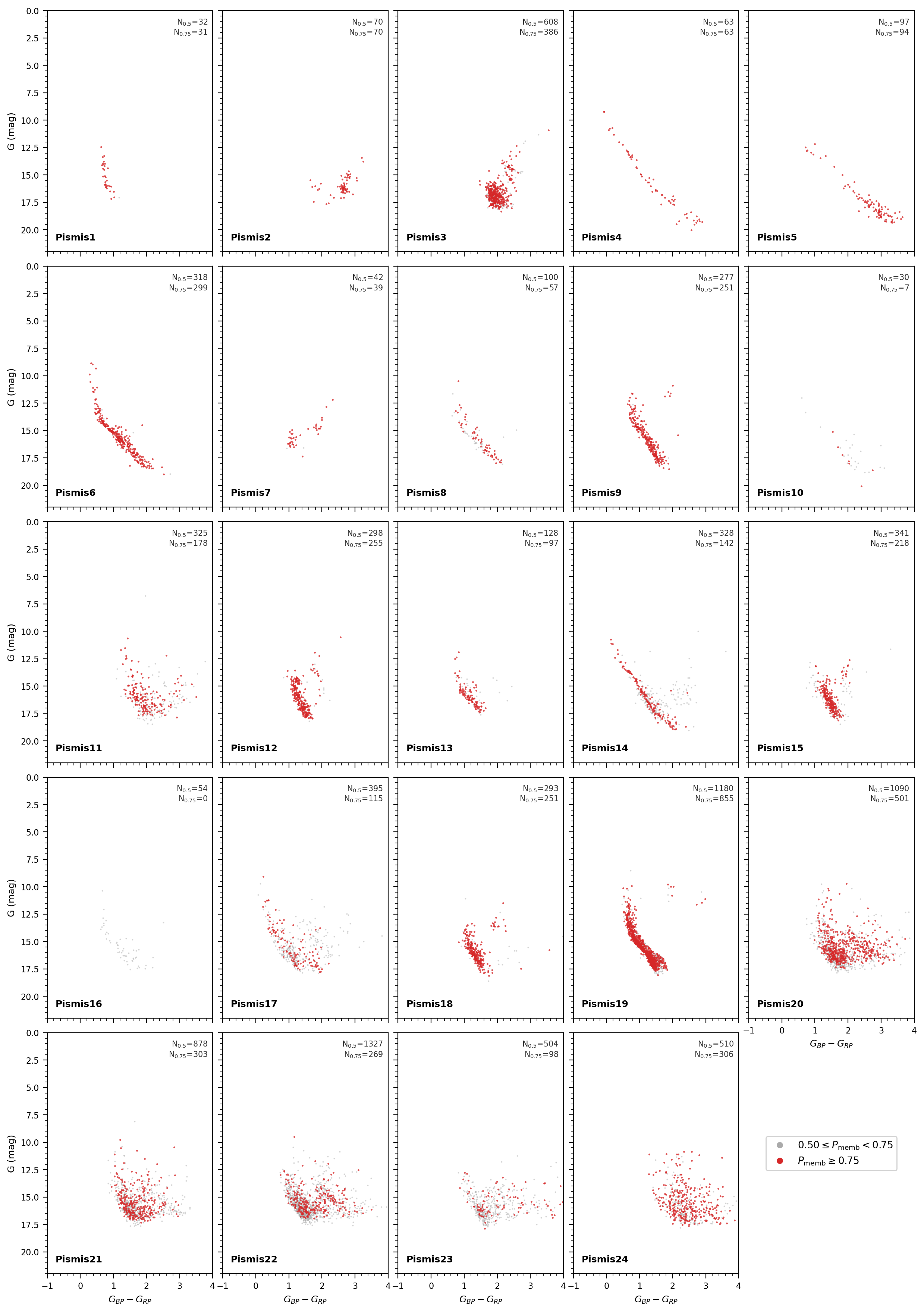}
    \caption{Color-Magnitude Diagrams (CMDs) for the initially surveyed Pismis clusters using \textit{Gaia} DR3 photometry. The $G$ vs. $(G_{\rm BP}-G_{\rm RP})$ planes illustrate the clear separation between identified cluster members (highlighted in red) and the surrounding field star population. This preliminary grid served as the basis for our selection process, allowing us to identify systems with statistically significant membership counts and well-defined stellar sequences suitable for high-precision isochrone fitting.}
    \label{fig:cmd_grid}
\end{figure*}

The original Pismis catalogue lists 24 open clusters. We evaluated the full list by applying the \textsc{UPMASK} membership algorithm detailed in Section~\ref{sec:membership} to the quality-filtered \textit{Gaia} DR3 data of Section~\ref{sec:query}; this algorithm assigns each star a membership probability $P$. As shown in the preliminary CMD grid (Figure~\ref{fig:cmd_grid}), several clusters were excluded from further analysis because they did not yield a sufficient number of members at a probability threshold of $P \geq 0.75$, or because their main sequences remained too heavily blended with field-star contamination for reliable isochrone fitting. This filtering left a final sample of 14 systems, selected on the basis of membership statistics and CMD morphology.

The specific causes for exclusion vary across the omitted systems. For instance, Pismis 16 yielded no member stars ($N_{0.75}=0$) at our high-probability threshold, while Pismis 10 provided too few candidates ($N_{0.75}=7$) to form a statistically significant sequence. Conversely, systems such as Pismis 20, 21, 22, and 23 displayed higher member counts but exhibited severe field star contamination that completely blended with the cluster main sequences. Pismis 24 presented a different observational challenge, as it is embedded within a dense nebulous region where high interstellar extinction obscures the cluster sequence on the CMD. Three further systems were removed for reasons specific to the fit rather than to the membership analysis. For Pismis~17 the \textsc{UPMASK} probability distribution failed to separate into cluster and field populations, remaining essentially flat between 0 and 1, and only $\sim$15 stars survived the quality cuts at the threshold used for the fits --- too few to constrain an isochrone. For Pismis~2 the reddening posterior converged on the upper boundary of its prior, $E(G_{\rm BP}-G_{\rm RP}) = 1.5$~mag, indicating that the extinction was not constrained by the data and that the associated age and distance could not be trusted. Pismis~11 lies in a crowded, heavily obscured field in which the recovered sequence could not be interpreted unambiguously. Together with the seven systems above, this accounts for all ten clusters of the original catalogue that are not analysed here. These contamination and extinction effects, visible as grey points in Figure~\ref{fig:cmd_grid}, introduce significant uncertainties that prevent the Bayesian Nested Sampling algorithm from reaching reliable convergence.

By excluding these poorly defined or heavily obscured systems, we focused the study on the 14 clusters that provide the most coherent stellar sequences. This selection restricts the analysis to the systems for which the member sequence can be isolated with confidence; the residual level of field contamination in each of the retained clusters is assessed quantitatively in Section~\ref{sec:fit_reliability}. The equatorial and Galactic coordinates adopted for each of the 14 selected clusters are listed in Table~\ref{tab:pismis_coordinates}.

\begin{table}
\centering
\caption{Equatorial (J2000) and Galactic coordinates of the selected Pismis clusters. Right ascension and declination are given in sexagesimal units, while Galactic coordinates are in degrees.}
\label{tab:pismis_coordinates}
\begin{tabular}{lcccc}
\hline
Cluster & RA (J2000) & Dec (J2000) & $l$ (deg) & $b$ (deg) \\
\hline

Pismis 1  & 08:18:18.8 & $-37$:06:06 & 255.1149 & -0.7100 \\
Pismis 3  & 08:31:20.4 & $-38$:38:16 & 257.8519 &  0.5049 \\
Pismis 4  & 08:35:09.6 & $-44$:25:26 & 262.9290 & -2.3630 \\
Pismis 5  & 08:37:37.1 & $-39$:35:23 & 259.3432 &  0.9119 \\
Pismis 6  & 08:39:07.9 & $-46$:13:36 & 264.7900 & -2.8929 \\
Pismis 7  & 08:41:08.2 & $-38$:41:58 & 259.0520 &  1.9951 \\
Pismis 8  & 08:41:36.2 & $-46$:16:01 & 265.0830 & -2.5789 \\
Pismis 9  & 08:42:56.2 & $-44$:55:55 & 264.1733 & -1.5739 \\
Pismis 12 & 09:20:01.4 & $-45$:07:57 & 268.6539 &  3.2080 \\
Pismis 13 & 09:22:07.0 & $-51$:06:06 & 273.1240 & -0.7650 \\
Pismis 14 & 09:29:52.8 & $-52$:46:48 & 275.1510 & -1.1390 \\
Pismis 15 & 09:34:44.7 & $-48$:02:36 & 272.4989 &  2.8611 \\
Pismis 18 & 13:36:54.8 & $-62$:05:29 & 308.2260 &  0.3120 \\
Pismis 19 & 14:30:40.0 & $-60$:53:21 & 314.7100 & -0.3070 \\

\hline
\end{tabular}
\end{table}

\section{Method}
\label{sec:methods}
\subsection{Membership Analyses and Astrometric Parameters}
\label{sec:membership}

Cluster membership determination relies on isolating the co-moving stellar population from the background and foreground field stars \citep{Castro-Ginard2022}. Because members of a single open cluster share a common space velocity and distance, they form distinct overdensities within the \textit{Gaia} astrometric space \citep{Cantat-Gaudin_2018}. We utilize these coherent vector groupings in proper motion ($\mu_{\alpha}\cos\delta, \mu_{\delta}$) and the localized distributions in trigonometric parallax ($\varpi$) as the primary multidimensional criteria to separate true cluster stars from field interlopers \citep{Monteiro_2021, Hunt_2023}.

In this study, we utilized \textit{Gaia}~DR3 proper motions and parallaxes to isolate cluster members from the field population across the 14 Pismis systems. Membership probabilities were calculated using the Unsupervised Photometric Membership Assignment in Stellar Clusters (\textsc{UPMASK}) algorithm \citep{Krone-Martins_2014} via the \texttt{pyUPMASK} Python package \citep{Pera2021}. The identified members form distinct overdensities in the Vector Point Diagrams (VPDs) and exhibit clear separation from the background field (see Table~A1).

The \textsc{UPMASK} algorithm operates through an iterative, non-parametric process to detect spatial and kinematic overdensities within a multidimensional parameter space. In each iteration, the five-dimensional astrometric data ($\alpha, \delta, \mu_{\alpha}\cos\delta, \mu_{\delta}, \varpi$) undergo Principal Component Analysis (PCA) to decorrelate and rescale the parameters. We deliberately restrict this feature space to astrometry alone, excluding photometric quantities such as $G$ and $(G_{\rm BP}-G_{\rm RP})$: since our subsequent isochrone fitting (Section~\ref{sec:method}) uses these same photometric observables as its primary constraint, selecting members along the same photometric axes would circularly bias the sample toward stars already close to an assumed isochrone, artificially suppressing genuine scatter and pre-main-sequence or evolved outliers. The algorithm then applies $k$-means clustering to partition the dataset into $k$ groups, where the hyperparameter $k$ scales with the expected cluster size. For internal validation, the spatial concentration of each identified group is compared against a uniform distribution generated via a random shuffling approach. Groups that exhibit a significantly higher concentration than the random distribution at a specified significance level are flagged as potential cluster members.

To account for observational uncertainties in the \textit{Gaia}~DR3 catalogue \citep{Gaia_DR3}, we performed the resampling process over 100 independent iterations. In each cycle, the astrometric parameters of every star were randomly drawn from a Gaussian distribution scaled by their measured values and associated uncertainties. The final membership probability ($P$) for each star corresponds to the frequency with which it was flagged as a member across all iterations. The membership probability enters the analysis at two distinct stages, and with two different thresholds. The first is the assessment of which clusters can be analysed at all (Section~\ref{sec:sampleselection}): there we count members at $P \geq 0.75$, a threshold low enough that a system failing it cannot be said to possess a recoverable sequence, and we use those counts --- together with the CMD morphology --- to reduce the original catalogue from 24 systems to 14. The second is the construction of the member samples that enter the isochrone fits themselves. For these we apply a uniform, more stringent cut of $P \geq 0.90$ to all 14 clusters, well above the $P > 0.5$ baseline common in the literature. The stricter cut is adopted deliberately: because every one of these systems lies close to the Galactic plane, sample purity rather than completeness is the limiting factor for isochrone fitting, and a uniform threshold keeps the 14 member catalogues mutually comparable, which a cluster-by-cluster optimisation would not. The member counts $N_{\rm mem}$ quoted throughout the paper and listed in Table~\ref{tab:ns_results} are therefore those at $P \geq 0.90$, whereas the counts used for the selection decision in Section~\ref{sec:sampleselection} are those at $P \geq 0.75$; the catalogue cross-matching of Section~\ref{subsec:comparison} uses a deliberately inclusive list ($P \geq 0.5$) so that the comparison is not biased by our own purity cut.

To maximize the purity of our final member list and mitigate the impact of unresolved binary systems, we applied a strict quality cut of $\text{RUWE} < 1.4$. This criterion effectively filters out sources with poorly constrained astrometric solutions, which are frequently associated with binary companions or spurious detections \citep{Lindegren_2018, Lindegren2021}. Adopting RUWE as a metric for astrometric stability ensures a clean sample of stars suitable for precise isochrone fitting without requiring arbitrary photometric adjustments. Additionally, the final membership was restricted to stars within the observationally determined limiting radius ($r_{\rm lim}^{\rm obs}$). The complete set of astrometric and derived fundamental parameters for all 14 Pismis clusters is provided in Table~\ref{tab:fundamental_results}.

\begin{table}
\centering
\caption{The adopted $k$ parameter for the \textsc{pyUPMASK} analysis and the resulting number of member stars ($N_{\rm mem}$) for the 14 Pismis clusters, at the uniform membership threshold $P \geq 0.90$ used for the isochrone fits (Section~\ref{sec:membership}).}
\label{tab:membership_results}
\begin{tabular}{lcc}
\hline\hline
Cluster Name & $k_{\rm means}$ & $N_{\rm mem}$ \\
\hline
Pismis 1  & 26  & 28  \\
Pismis 3  & 11  & 120 \\
Pismis 4  & 26  & 62  \\
Pismis 5  & 10  & 90  \\
Pismis 6  & 21  & 259 \\
Pismis 7  & 11  & 38  \\
Pismis 8  & 21  & 49  \\
Pismis 9  & 11  & 193 \\
Pismis 12 & 11  & 219 \\
Pismis 13 & 8   & 86  \\
Pismis 14 & 12  & 88  \\
Pismis 15 & 16  & 182 \\
Pismis 18 & 16  & 232 \\
Pismis 19 & 9   & 571 \\
\hline
\end{tabular}
\begin{flushleft}
\end{flushleft}
\end{table}

\subsection{Age Determination}
\label{sec:method}

Theoretical isochrones were computed using the PARSEC stellar evolution models \citep{Bressan_2012} accessed through the CMD~3.9 web interface\footnote{\url{http://stev.oapd.inaf.it/cmd}}, adopting the Gaia~DR3 photometric system \citep{Gaia_DR3}. The grid spans $\log(\mathrm{Age/yr}) \in [6.0,\,10.1]$ in steps of 0.05\,dex (72 values), denoted $\log(\mathrm{Age})$ throughout the remainder of this paper for brevity, and initial metallicity $Z_{\rm ini}$ over 26 values distributed log-linearly between $Z = 0.0001$ ($[\mathrm{Fe/H}]\approx-2.18$) and $Z = 0.0500$ ($[\mathrm{Fe/H}]\approx+0.52$), with the solar value $Z_{\odot}=0.0152$ explicitly included.

To enable rapid likelihood evaluations, we trained a \textsc{scikit-learn} \texttt{DecisionTreeRegressor} as a high-dimensional interpolator on the grid, mapping the input triplet $(\log(\mathrm{Age}),\, Z_{\rm ini},\, M_{\rm ini})$ to the observed Gaia quantities $(G,\, G_{\rm BP}-G_{\rm RP})$. Photometric extinction was applied following \citet{Wang2019}, using the coefficients $A_G / E(G_{\rm BP}-G_{\rm RP}) = 1.890$ and $E(G_{\rm BP}-G_{\rm RP}) / E(B{-}V) = 1.305$.

We infer four global cluster parameters $\boldsymbol{\theta} = (\log(\mathrm{Age}),\, Z_{\rm ini},\, d_{\rm pc},\, E(G_{\rm BP}-G_{\rm RP}))$ via Dynamic Nested Sampling \citep{Higson2019} using the \textsc{dynesty} package \citep{Speagle2020}. The sampling was performed with 200 live points, employing a \texttt{multi}-ellipsoid bounding strategy and random-walk proposals (\texttt{rwalk}). Convergence was strictly defined by the evidence tolerance criterion $\Delta\ln\mathcal{Z} \leq 0.01$, where $\mathcal{Z}$ denotes the Bayesian evidence and is not to be confused with the initial metallicity $Z_{\rm ini}$ introduced above. Earlier versions of this isochrone-fitting framework have been applied in previous open-cluster studies \citep{Tanik2025, Karagoz_2025, Yontan2026}.

For each proposed parameter set $\boldsymbol{\theta}$, a synthetic colour--magnitude diagram (CMD) is generated from the trained interpolator. Each observed star $i$ is matched to its closest counterpart on the theoretical isochrone in the $(G_{\rm BP}-G_{\rm RP},\,G)$ plane. Denoting by $\sigma_{c,i}$ and $\sigma_{G,i}$ the observed photometric uncertainties in colour and magnitude, respectively, we define the squared distance $\chi^2_{i}$ as:

\begin{widetext}
\begin{equation}
    \chi^2_{i}(\boldsymbol{\theta}) = \frac{[(G_{\rm BP}-G_{\rm RP})_{obs,i} - (G_{\rm BP}-G_{\rm RP})_{mod}]^2}{\sigma_{c,i}^2} + \frac{[G_{obs,i} - G_{mod}]^2}{\sigma_{G,i}^2},
    \label{eq:chi2_def}
\end{equation}
\end{widetext}

\noindent To account for unmodelled systematics, potential binary sequences, and small-scale imperfections in stellar models, these uncertainties were inflated by a factor of 2.5. The final membership-weighted, truncated log-likelihood is given by:

\begin{equation}
\ln\mathcal{L}(\boldsymbol{\theta}) =
  -\frac{1}{2}\sum_{i} w_i
  \left[\min\!\left(\chi^2_{i},\,\chi^2_{\rm max}\right)
  + \ln\!\left(2\pi\,\sigma_{c,i}\,\sigma_{G,i}\right)\right],
\label{eq:loglike}
\end{equation}

\noindent where $w_i = P_{{\rm mem},i}/\sum_j P_{{\rm mem},j}$ is the normalised UPMASK membership probability of star $i$, and $\chi^2_{\rm max}=50$ is a robustness threshold applied to prevent runaway contributions from outliers or remaining field contaminants. The weighting was applied exclusively to member stars that exceeded the selected probability threshold ($P \geq P_{\rm threshold}$).

Each of the four free parameters is assigned a physically motivated prior as follows:

\paragraph{Age:} 
In the absence of strong prior constraints on the age of each cluster, we adopt a non-informative uniform prior over the full range of the isochrone grid: $\pi(\log(\mathrm{Age})) = \mathcal{U}(6.0,\,10.0)$. This ensures the posterior is driven primarily by the CMD morphology.

\paragraph{Metallicity:} 
Given the lack of spectroscopic data for most Pismis clusters, we model the metallicity prior based on the distribution of solar-neighbourhood open clusters \citep{Netopil_2016}. We adopt a skew-normal distribution:
\begin{equation}
  \pi([\mathrm{Fe/H}]) \propto \mathrm{SN}(\mu=0.0,\,\sigma=0.3,\,\alpha=-2),
\end{equation}
which accounts for the observed asymmetry and the preference for near-solar values. The sampled $[\mathrm{Fe/H}]$ is converted to $Z_{\rm ini}$ using $Z = Z_{\odot}\,10^{[\mathrm{Fe/H}]}$.

\paragraph{Distance:} 
We utilize Gaia~DR3 parallaxes to anchor the distance scale. For stars satisfying quality cuts ($\varpi > 0$, $\sigma_\varpi/|\varpi| \leq 0.20$, $\mathrm{RUWE} \leq 1.4$), we adopt a truncated Gaussian prior:
\begin{equation}
  \pi(d) \propto \mathcal{N}(\tilde{d},\,15^2) \;\Big|_{d\,>\,0},
\end{equation}
where $\tilde{d}$ is the median distance derived from the cluster members. This informative, cluster-specific prior differs in form from the exponentially-decreasing space-density prior of \citet{Bailer-Jones21}, which is designed for individual field stars with a single, often noisy parallax measurement and therefore requires a broad, direction-dependent length-scale prior to regularise the inversion. In our case, the prior mean $\tilde{d}$ is itself the median of tens to hundreds of parallaxes from co-distant cluster members, which already reduces the per-star parallax uncertainty by a factor of order $\sqrt{N_{\rm mem}}$; the narrow 15\,pc width reflects the residual uncertainty in this already well-constrained mean rather than an assumed astrophysical distance distribution. The two approaches are therefore consistent in spirit -- both regularise the parallax-to-distance inversion with a prior appropriate to the information at hand -- but ours is tailored to the more constrained problem of a resolved, multi-member cluster rather than a single field star. We note that neither $\tilde{d}$ nor the individual member parallaxes feeding into it were corrected for the Gaia DR3 parallax zero-point offset (e.g., \citealt{Lindegren2021}) or reprocessed through the \citet{Bailer-Jones21} distance-estimation pipeline; the quoted distances and their uncertainties therefore reflect only the internal statistical precision of the Bayesian posterior and do not explicitly propagate this external systematic. Given the typical magnitude of the zero-point offset ($\sim$10--40\,$\mu$as, corresponding to a few percent in distance at the distances of our sample) relative to the 15\,pc prior width and the posterior uncertainties reported in Table~\ref{tab:ns_results}, we expect this to be a subdominant, though not negligible, source of additional systematic uncertainty, particularly for the most distant clusters in the sample (e.g., Pismis~1 and~7, at $\sim$5.4\,kpc).

\paragraph{Reddening:} 
The reddening $E(G_{\rm BP}-G_{\rm RP})$ is assigned a broad uniform prior $\pi(E) = \mathcal{U}(0.0,\,5.0)$. To optimize the initial state of the nested sampler, we provide a starting estimate $\hat{E}_{\rm SFD}$ using \citet{Schlegel1998} dust maps via \textsc{mwdust} \citep{mwdust}, corrected for a Galactic scale height of $h = 125$\,pc.

\begin{table*}
  \centering
  \caption{Bayesian nested-sampling parameters for 14 Pismis open clusters (PARSEC isochrones). Columns list: number of member stars after quality cuts ($N_{\rm mem}$), logarithmic age, age in Gyr, initial metallicity $Z_{\rm ini}$, iron abundance [Fe/H], heliocentric distance $d$, and reddening $E(G_{\rm BP}-G_{\rm RP})$. All values are posterior medians; uncertainties are 16th/84th percentile intervals.}
  \label{tab:ns_results}
  \begin{tabular}{lccccccc}
    \hline\hline
    Cluster & $N_{\rm mem}$ & $\log(\mathrm{Age})$ & Age (Gyr) & $Z_{\rm ini}$ & ${[\rm Fe/H]}$ (dex) & $d$ (pc) & $E(G_{\rm BP}{-}G_{\rm RP})$ \\
    \hline
    Pismis~1 & 28 & $7.75^{+0.20}_{-0.21}$ & $0.056^{+0.032}_{-0.021}$ & $0.0092^{+0.0045}_{-0.0033}$ & $-0.22^{+0.17}_{-0.19}$ & $5445^{+14}_{-15}$ & $0.955^{+0.040}_{-0.045}$ \\
    Pismis~3 & 120 & $9.04^{+0.18}_{-0.21}$ & $1.104^{+0.553}_{-0.430}$ & $0.0111^{+0.0066}_{-0.0043}$ & $-0.14^{+0.20}_{-0.21}$ & $2293^{+15}_{-15}$ & $1.575^{+0.142}_{-0.143}$ \\
    Pismis~4 & 62 & $8.35^{+0.55}_{-0.50}$ & $0.224^{+0.580}_{-0.154}$ & $0.0084^{+0.0047}_{-0.0033}$ & $-0.26^{+0.19}_{-0.21}$ & $690^{+15}_{-15}$ & $0.265^{+0.246}_{-0.143}$ \\
    Pismis~5 & 90 & $7.04^{+0.50}_{-0.48}$ & $0.011^{+0.024}_{-0.007}$ & $0.0098^{+0.0052}_{-0.0038}$ & $-0.19^{+0.19}_{-0.22}$ & $926^{+14}_{-15}$ & $0.600^{+0.253}_{-0.250}$ \\
    Pismis~6 & 259 & $7.88^{+0.41}_{-0.55}$ & $0.076^{+0.117}_{-0.054}$ & $0.0100^{+0.0050}_{-0.0037}$ & $-0.18^{+0.18}_{-0.20}$ & $1815^{+15}_{-15}$ & $0.736^{+0.162}_{-0.119}$ \\
    Pismis~7 & 38 & $8.64^{+0.32}_{-0.11}$ & $0.433^{+0.464}_{-0.095}$ & $0.0092^{+0.0054}_{-0.0034}$ & $-0.22^{+0.20}_{-0.20}$ & $5367^{+15}_{-15}$ & $1.075^{+0.085}_{-0.211}$ \\
    Pismis~8 & 49 & $8.28^{+0.43}_{-0.53}$ & $0.191^{+0.322}_{-0.135}$ & $0.0108^{+0.0056}_{-0.0043}$ & $-0.15^{+0.18}_{-0.22}$ & $1865^{+16}_{-15}$ & $0.981^{+0.167}_{-0.148}$ \\
    Pismis~9 & 193 & $8.30^{+0.32}_{-0.46}$ & $0.200^{+0.216}_{-0.131}$ & $0.0111^{+0.0064}_{-0.0044}$ & $-0.14^{+0.20}_{-0.22}$ & $1995^{+16}_{-14}$ & $0.928^{+0.134}_{-0.136}$ \\
    Pismis~12 & 219 & $9.18^{+0.30}_{-0.17}$ & $1.518^{+1.505}_{-0.492}$ & $0.0104^{+0.0049}_{-0.0036}$ & $-0.17^{+0.17}_{-0.19}$ & $2233^{+15}_{-15}$ & $0.798^{+0.146}_{-0.126}$ \\
    Pismis~13 & 86 & $8.35^{+0.26}_{-0.30}$ & $0.221^{+0.186}_{-0.110}$ & $0.0176^{+0.0052}_{-0.0061}$ & $+0.06^{+0.11}_{-0.19}$ & $2900^{+15}_{-15}$ & $0.919^{+0.091}_{-0.114}$ \\
    Pismis~14 & 88 & $8.78^{+0.36}_{-0.39}$ & $0.606^{+0.775}_{-0.357}$ & $0.0097^{+0.0048}_{-0.0039}$ & $-0.20^{+0.17}_{-0.22}$ & $1291^{+15}_{-14}$ & $0.442^{+0.119}_{-0.107}$ \\
    Pismis~15 & 182 & $9.09^{+0.35}_{-0.21}$ & $1.239^{+1.513}_{-0.467}$ & $0.0124^{+0.0062}_{-0.0044}$ & $-0.09^{+0.18}_{-0.19}$ & $2421^{+14}_{-16}$ & $0.969^{+0.155}_{-0.165}$ \\
    Pismis~18 & 232 & $8.54^{+0.32}_{-0.43}$ & $0.346^{+0.383}_{-0.217}$ & $0.0143^{+0.0060}_{-0.0060}$ & $-0.03^{+0.15}_{-0.24}$ & $2807^{+15}_{-14}$ & $1.160^{+0.135}_{-0.187}$ \\
    Pismis~19 & 571 & $7.69^{+0.52}_{-0.55}$ & $0.049^{+0.114}_{-0.035}$ & $0.0103^{+0.0060}_{-0.0041}$ & $-0.17^{+0.20}_{-0.22}$ & $2400^{+14}_{-15}$ & $0.799^{+0.189}_{-0.340}$ \\
    \hline
  \end{tabular}
\end{table*}

\begin{table*}
\centering
\caption{Spectroscopic statistics of the Pismis open clusters included in the study after applying the membership probability threshold. $N_{\rm mem}$ denotes the total number of stars exceeding the threshold; $N_{\rm spec}$ represents the number of stars with available spectroscopic measurements. Data sources: Gaia--ESO DR5.1, MWM19 (SDSS-V) \citep{Kollmeier2026}, and GALAH DR4 \citep{Buder2025}.}
\label{tab:pismis_stats}
\setlength{\tabcolsep}{6pt}
\begin{tabular}{lcccccccc}
\hline\hline
Cluster & Source & $N_{\rm spec}$ &
$\tilde{v}_{\rm r}$ (km~s$^{-1}$) & $\sigma_{v_{\rm r}}$ (km~s$^{-1}$) &
${[\rm Fe/H]}$ (dex) & $\sigma_{\rm [Fe/H]}$ \\
\hline
Pismis1 &  MWM19 & 6 & +97.2 & 339.5 & -0.39 & 0.86 \\
Pismis3 &  ... & ... & ... & ... & ... & ... \\
Pismis4 &  ... & ... & ... & ... & ... & ... \\
Pismis5 &  ... & ... & ... & ... & ... & ... \\
Pismis6 & ... & ... & ... & ... & ... & ... \\
Pismis7 &  ... & ... & ... & ... & ... & ... \\
Pismis8 &  ... & ... & ... & ... & ... & ... \\
Pismis9 &  ... & ... & ... & ... & ... & ... \\
Pismis12 & ... & ... & ... & ... & ... & ... \\
Pismis13 &  ... & ... & ... & ... & ... & ... \\
Pismis14 &  GALAH DR4 & 2 & +8.3 & 5.2 & -0.39 & 0.10 \\
Pismis15 &  Gaia--ESO & 70 & +34.6 & 7.3 & -0.14 & 0.23 \\
Pismis18 &  Gaia--ESO & 25 & -27.6 & 7.0 & -0.09 & 0.21 \\
Pismis19 & GALAH DR4 & 4 & -29.5 & 131.5 & -0.23 & 0.57 \\
\hline
\end{tabular}
\end{table*}
\begin{table*}
\centering
\caption{Gaia XP stellar parameters for Pismis open clusters matched to the
JDrgb catalogue \citep{Andrae2023}.
$N_{\rm XP}$: stars matched in JDrgb.
Columns give median total metallicity $\widetilde{[\rm M/H]}$,
median alpha enhancement $\widetilde{[\alpha/\rm M]}$ \citep{Li2024},
median Gaia radial velocity $\tilde{v}_{\rm r}$, its dispersion
$\sigma_{v_{\rm r}}$, and median formal RV uncertainty $\tilde{\epsilon}_{v_{\rm r}}$.}
\label{tab:pismis_gaiaxp}
\setlength{\tabcolsep}{6pt}

\begin{tabular}{lcc rrrrrrr}
\hline\hline
Cluster & $N_{\rm XP}$ &
$\widetilde{[\rm M/H]}$ & $\sigma_{\rm [M/H]}$ &
$\widetilde{[\alpha/\rm M]}$ & $\sigma_{\alpha/\rm M}$ &
$\tilde{v}_{\rm r}$ & $\sigma_{v_{\rm r}}$ & $\tilde{\epsilon}_{v_{\rm r}}$ \\
 & & & & & & \multicolumn{3}{c}{(km~s$^{-1}$)} \\
\hline
Pismis1  & 4  & -0.36 & 0.37 &  0.11 & 0.06 &  83.1 & 32.4 & 1.70 \\
Pismis3  & 31 & -0.08 & 0.15 &  0.03 & 0.03 &  30.5 &  6.7 & 1.89 \\
Pismis4  & 0  & \ldots & \ldots & \ldots & \ldots & \ldots & \ldots & \ldots \\
Pismis5  & 1  &  0.31 & \ldots &  0.00 & \ldots &  39.3 & \ldots & 0.37 \\
Pismis6  & 5  & -0.14 & 0.31 &  0.09 & 0.08 &  51.9 & 29.3 & 1.71 \\
Pismis7  & 16 & -0.19 & 0.12 &  0.02 & 0.03 &  74.3 &  7.8 & 3.74 \\
Pismis8  & 0  & \ldots & \ldots & \ldots & \ldots & \ldots & \ldots & \ldots \\
Pismis9  & 10 & -0.01 & 0.15 & -0.02 & 0.05 &  25.8 & 11.4 & 0.28 \\
Pismis12 & 13 & -0.09 & 0.13 &  0.01 & 0.04 &  32.7 & 13.4 & 1.52 \\
Pismis13 & 0  & \ldots & \ldots & \ldots & \ldots & \ldots & \ldots & \ldots \\
Pismis14 & 3  & -0.29 & 0.06 &  0.05 & 0.02 &  51.3 & 12.9 & 1.67 \\
Pismis15 & 16 & -0.07 & 0.10 &  0.01 & 0.02 &  34.8 & 11.7 & 1.57 \\
Pismis18 & 21 &  0.16 & 0.06 & -0.02 & 0.02 & -28.5 &  4.3 & 1.26 \\
Pismis19 & 34 &  0.19 & 0.18 &  0.00 & 0.04 & -24.9 & 31.5 & 1.90 \\
\hline
\end{tabular}
\end{table*}

\section{Results}
\label{sec:results}
\subsection{Derived Astrophysical Parameters and Statistical Diagnostics for Pismis 18}

\begin{figure*}
    \centering
    \includegraphics[width=0.85\textwidth]{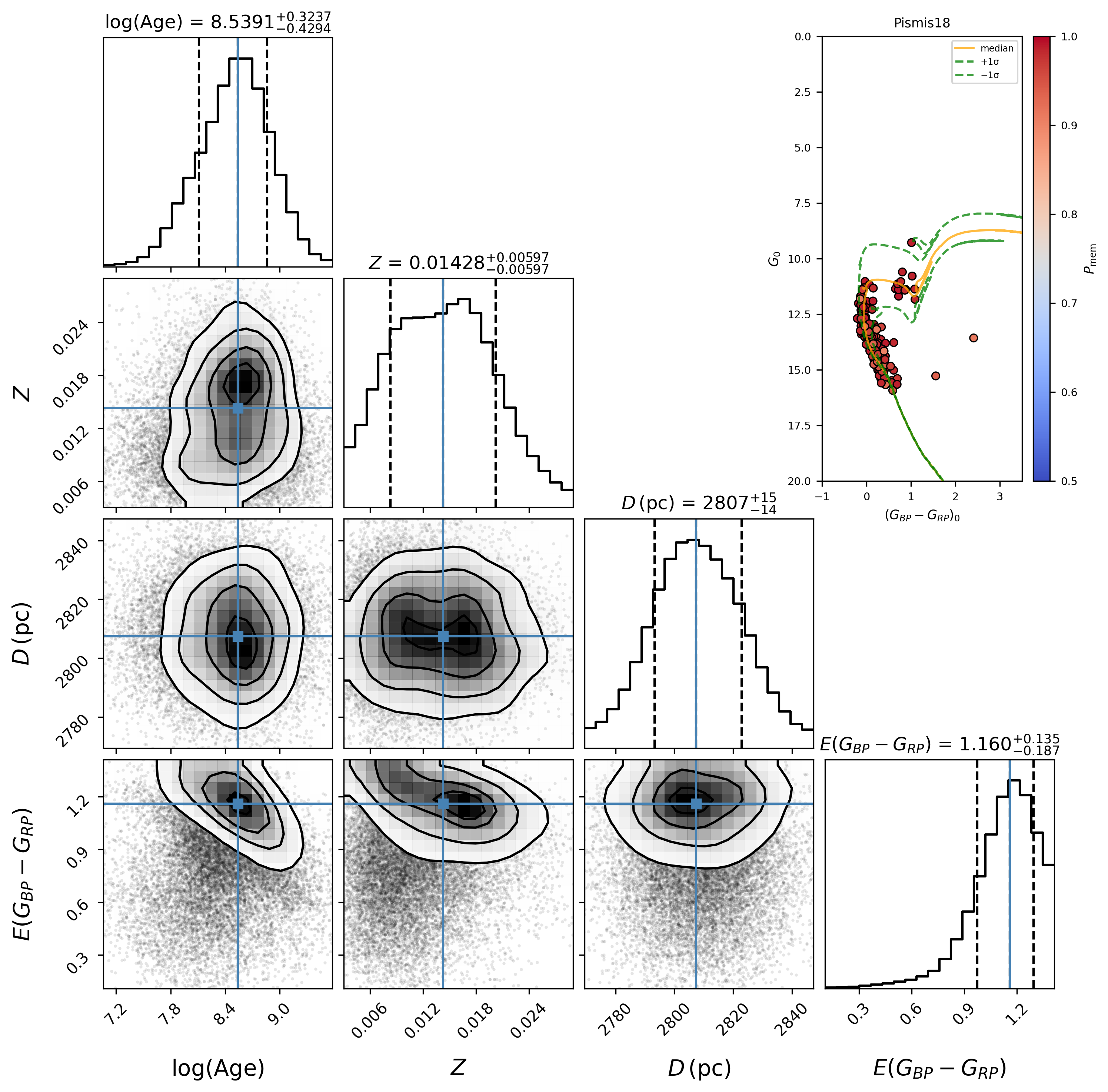}
    \caption{Bayesian parameter estimation results for the representative cluster Pismis 18. The projection matrix displays the one- and two-dimensional posterior probability distributions for the logarithmic age ($\log(\mathrm{Age})$), metallicity ($Z$), heliocentric distance ($D$), and interstellar reddening ($E(G_{\rm BP} - G_{\rm RP})$). In the 2D covariance contours, the shaded regions represent the $1\sigma, 2\sigma,$ and $3\sigma$ confidence levels, respectively. The top-right panel illustrates the $G$ vs. $(G_{\rm BP}-G_{\rm RP})$ colour-magnitude diagram (CMD), with stars colour-coded according to their \textsc{UPMASK} membership probabilities. The best-fit PARSEC isochrone is denoted by a solid orange line, while green dashed lines represent the $1\sigma$ uncertainty envelope derived from the posterior samples.}
    \label{fig:pismis18_corner}
\end{figure*}

In this section, we present the fundamental parameters derived for our cluster sample. Pismis~18 is treated as a representative case, while the diagnostic plots and parameters for the remaining 13 clusters are provided as online supplementary material (see Data Availability). Figure~\ref{fig:pismis18_corner} displays the posterior distributions alongside the corresponding isochrone fit. As shown in the Colour-Magnitude Diagram (top-right panel), stars with high membership probabilities ($P \geq 0.90$) define a well-defined Main Sequence. The median isochrone (orange solid line) and the $1\sigma$ uncertainty envelope (green dashed lines) are highly consistent with the observed data, particularly near the turn-off point. We determine the cluster age to be $\log(\mathrm{Age}) = 8.54^{+0.32}_{-0.43}$ ($\sim 346$~Myr).

\subsubsection{Posterior Distributions and Parameter Degeneracies}
The multidimensional parameter space---comprising age, metallicity ($Z$), distance ($d_{\rm pc}$), and interstellar reddening ($E(G_{\rm BP} - G_{\rm RP})$)---reveals the standard interdependencies inherent to isochrone fitting. The projected two-dimensional posterior contours in the corner plot display the structure of the likelihood surface for Pismis~18. A clear anti-correlation between $\log(\mathrm{Age})$ and $E(G_{\rm BP} - G_{\rm RP})$ is evident, reflecting the well-known degeneracy between cluster age and interstellar extinction. Despite this coupling, the Nested Sampling algorithm provides well-constrained posterior probability distributions, resolving the high-probability parameter space without introducing significant systematic offsets in the final estimates.

This results in a well-defined peak for the cluster age, indicating that for Pismis~18 the likelihood constrains age and reddening separately despite their correlation; we show in Section~\ref{sec:fit_reliability} that this separation is less effective for the differentially-reddened systems of the sample. Regarding the spatial parameters, our analysis yields a heliocentric distance of $d = 2807^{+15}_{-14}$~pc. The symmetric, near-Gaussian profile of the distance posterior reflects the direct constraint provided by the \textit{Gaia}~DR3 parallaxes at kiloparsec scales. Furthermore, although the metallicity ($Z \approx 0.014$) was initially constrained by a skew-normal prior, the final posterior is shaped by the photometric likelihood; however, the metallicity constraints remain partially prior-dependent due to the limited sensitivity of \textit{Gaia} broadband photometry to abundance variations. This convergence aligns Pismis~18 with the expected chemical properties of the Galactic disc. The parameters derived for Pismis~18 demonstrate the applicability of this framework to open clusters across diverse Galactocentric environments. Detailed diagnostic outputs for the remaining 13 clusters in our sample are documented as online supplementary material (see Data Availability).

\subsection{Comparison with Previous Catalogues}
\label{subsec:comparison}
\subsubsection{Comparison of Membership Samples}

\begin{figure*}
    \centering
    \includegraphics[width=\textwidth]{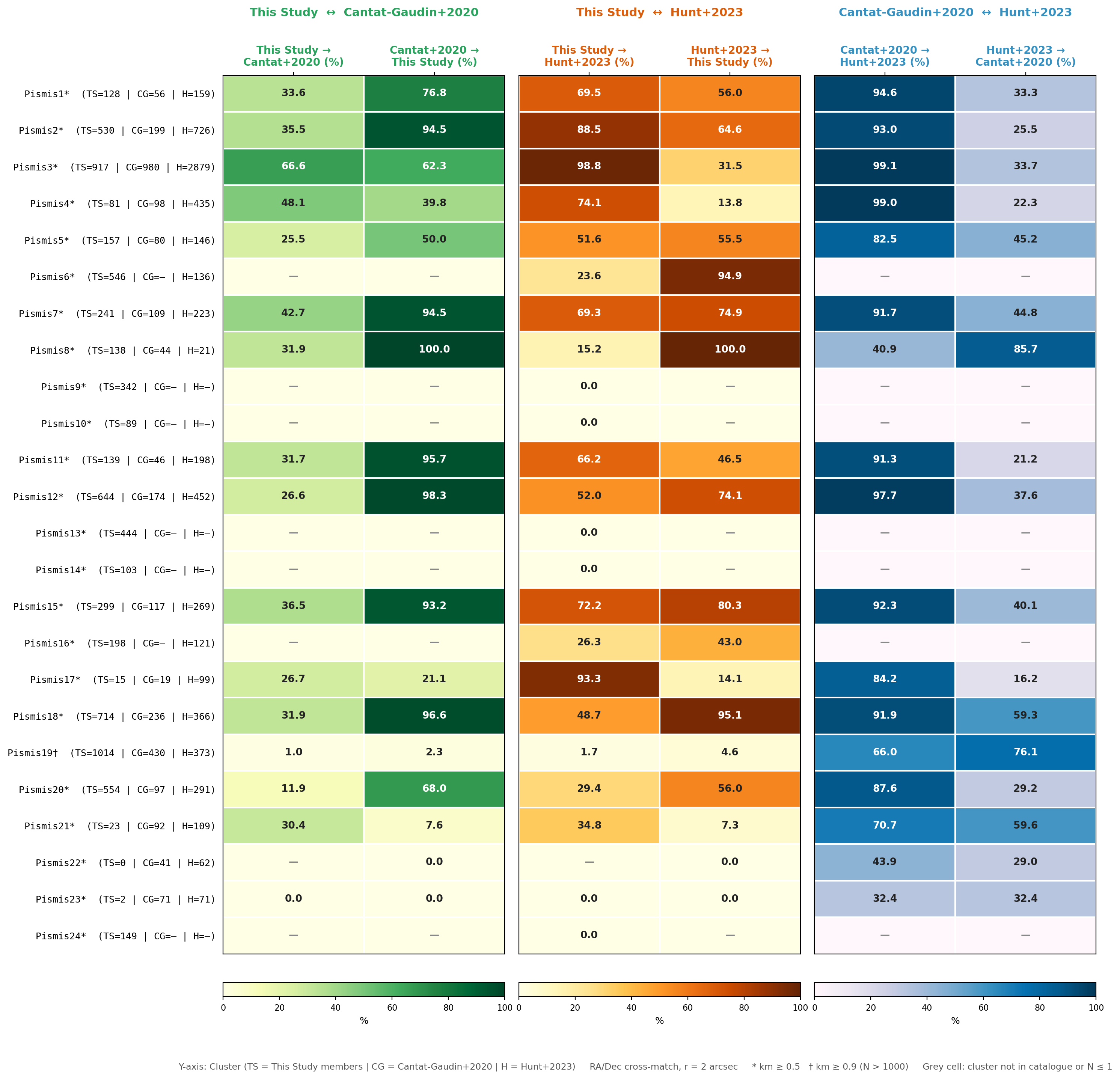}
    \caption{Membership comparison matrix across this work and established catalogues for the 24 Pismis clusters. Each cell provides the cross-match count (top), the Jaccard index ($J = |A \cap B| / |A \cup B| \times 100$, color-coded), and the percentage of star recovery between samples (bottom). The significant population of unique members identified in this study highlights the efficacy of cluster-specific analysis in revealing stellar populations previously missed by large-scale automated surveys.}
    \label{fig:comparison_matrix}
\end{figure*}

We performed a cross-match between the members identified in this study and the established catalogues of \citet{Cantat-Gaudin_2020} (CG20) and \citet{Hunt_2023} (H23) using a matching radius of $r = 2$ arcsec. The resulting comparison matrix (Figure~\ref{fig:comparison_matrix}) provides the star recovery percentages and the Jaccard index, defined as $J = |A \cap B| / |A \cup B|$, i.e.\ the size of the intersection between two membership sets $A$ and $B$ divided by the size of their union, for each system.

The most notable feature of the comparison is the high number of unique members recovered in this study, particularly for clusters like Pismis~18 and Pismis~19. While our sample includes a significant majority of stars found in CG20 (with recovery rates reaching or exceeding 70\% in six of the ten clusters for which CG20 provides a counterpart, and surpassing 90\% in Pismis~7, 8, 12, 15, and 18), our total counts (TS) are significantly higher. The effect is most extreme for Pismis~19, the most crowded field in the sample, where the overlap with CG20 falls to a few per cent of our member list. This discrepancy indicates that global automated pipelines likely omit a substantial portion of the cluster population in dense Galactic fields.

Additionally, Figure~\ref{fig:comparison_matrix} shows low consistency between CG20 and H23 for several clusters, such as Pismis 8 and Pismis 19. In the case of Pismis 22 and 23, the recovery rate between our study and previous catalogues is effectively zero. This lack of overlap supports our decision to exclude these systems from the final astrophysical parameter estimation, as the high field contamination prevents reliable membership assignment even at the $P \geq 0.75$ level. These results demonstrate that cluster-specific analysis is required to obtain a complete census of the member population, especially for systems located in crowded regions.

\subsection{Astrophysical Parameters and Bayesian Uncertainty Analysis}
\label{subsec:params}

\subsubsection{Comparison of Astrophysical Parameters}

\begin{figure*}
    \centering
    \includegraphics[width=0.9\textwidth]{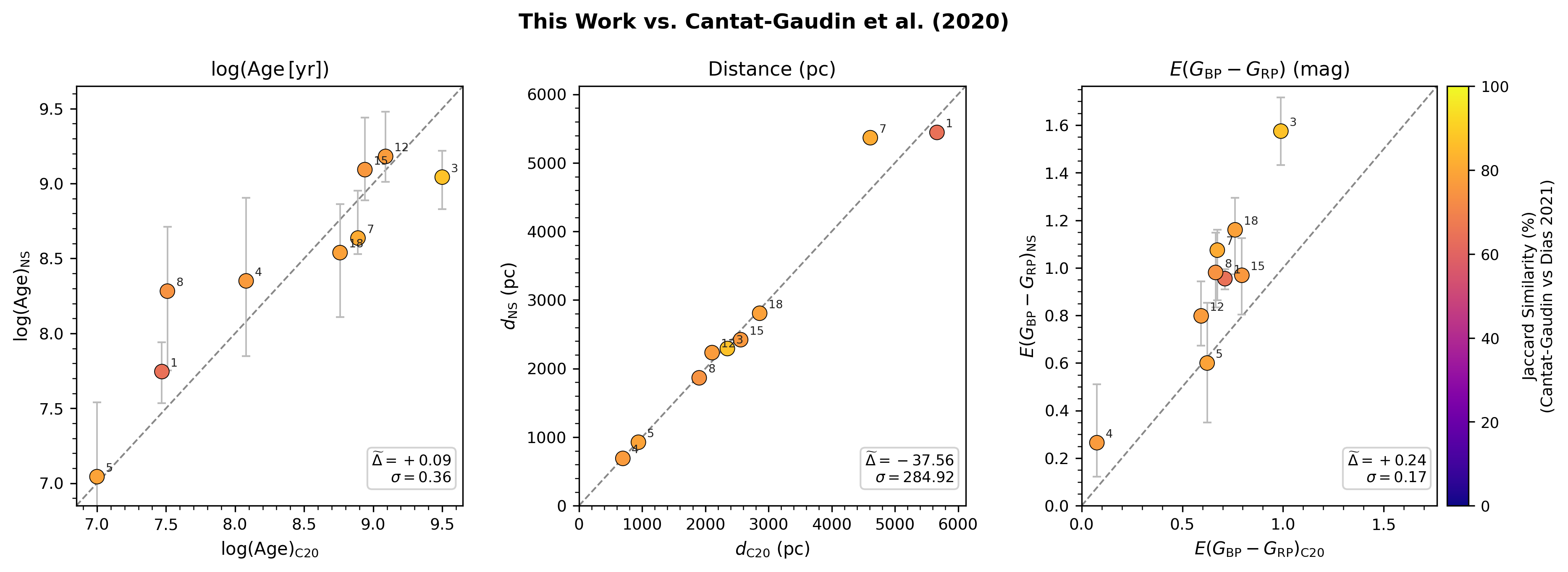}
    \caption{Comparison of fundamental parameters derived in this work with those from \citet{Cantat-Gaudin_2020}. Solid lines represent the 1:1 identity. Mean offsets ($\Delta$) and standard deviations ($\sigma$) are indicated in each panel.}
    \label{fig:comp_cantat}
\end{figure*}

We compared the fundamental parameters derived in this study---$\log(\mathrm{Age})$, distance, and reddening---with the values from \citet{Cantat-Gaudin_2020} for the 14 clusters in common. Figure~\ref{fig:comp_cantat} shows these comparisons, where points are color-coded according to the Jaccard similarity index. The distance estimates are consistent with the literature, showing a mean offset of $\bar{\Delta} = -37.56$~pc and a standard deviation of $\sigma = 284.92$~pc. 

The reddening $E(G_{\rm BP}-G_{\rm RP})$ shows a systematic offset of $+0.24$~mag relative to the \citet{Cantat-Gaudin_2020} values. Similarly, our age estimates are systematically older by a mean of $\bar{\Delta} = +0.09$~dex ($\sigma = 0.36$). The scatter in the age and reddening panels correlates with the membership differences; clusters with lower Jaccard similarity indices exhibit larger deviations from the 1:1 identity line. These systematic shifts indicate that the inclusion of the unique member populations identified via our $P \geq 0.90$ threshold directly influences the results of the isochrone fitting process compared to automated global pipelines.

The astrophysical parameters for the 14 Pismis clusters, determined through Bayesian nested sampling, are presented in Table~\ref{tab:ns_results}. Our approach provides not only the posterior medians but also the 16th and 84th percentile intervals, offering a rigorous assessment of the uncertainties associated with each parameter.

\begin{figure*}
    \centering
    \includegraphics[width=0.9\textwidth]{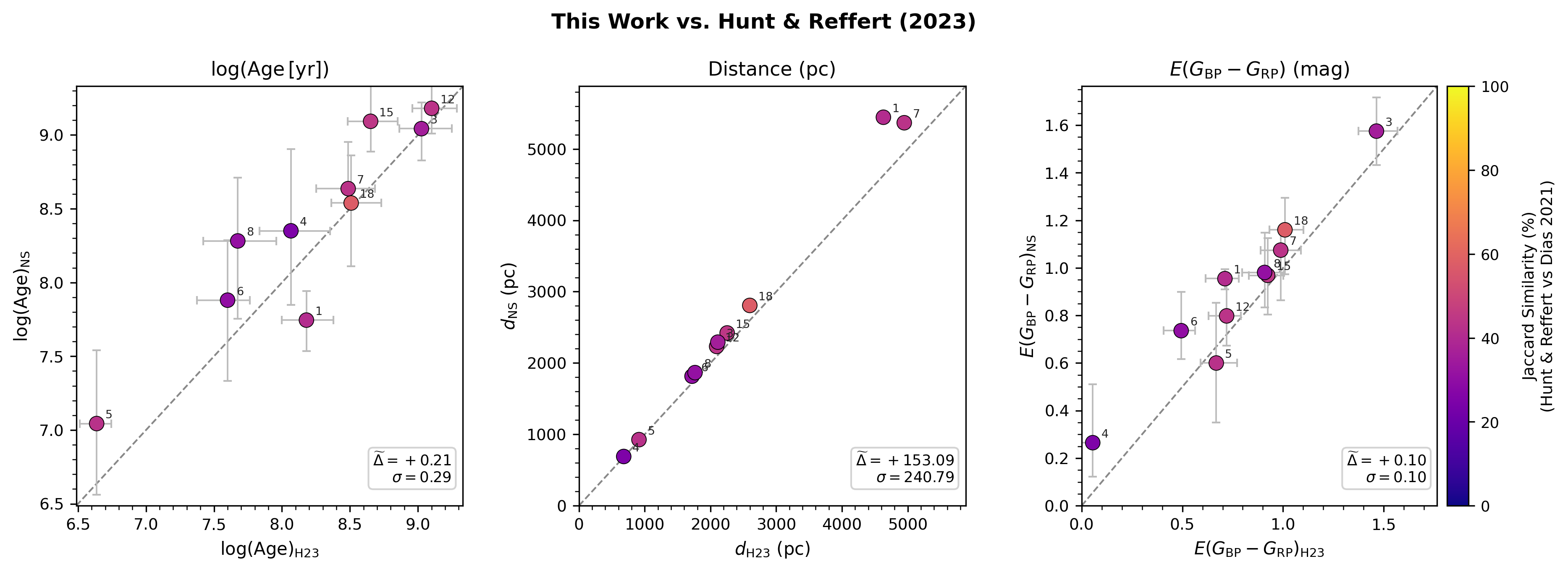}
    \caption{Same as Figure \ref{fig:comp_cantat}, but for comparison with the \citet{Hunt_2023} catalogue.}
    \label{fig:comp_hunt}
\end{figure*}

A similar comparison was performed with the results of \citet{Hunt_2023} (H23), as shown in Figure \ref{fig:comp_hunt}. The distance estimates relative to H23 show a mean offset of $\bar{\Delta} = +153.09$~pc ($\sigma = 240.79$~pc). Our age estimates are also systematically older by $\bar{\Delta} = +0.21$~dex ($\sigma = 0.29$), while the reddening shows a systematic shift of $+0.10$~mag ($\sigma = 0.10$). 

As indicated by the color-coding in Figure \ref{fig:comp_hunt}, the Jaccard similarity between this study and H23 is generally lower than that observed with CG20. The lower consistency in membership lists is evident in the age and distance panels, where clusters with the lowest similarity indices exhibit the largest deviations from the 1:1 identity line. These differences arise from the contrasting membership selection criteria and the choice of isochrone models between the two studies. The recovery of unique members in our analysis ($P \geq 0.90$) contributes to the observed offsets, confirming that astrophysical parameters are sensitive to the initial stellar population census in crowded Galactic regions.

\begin{figure*}
    \centering
    \includegraphics[width=0.9\textwidth]{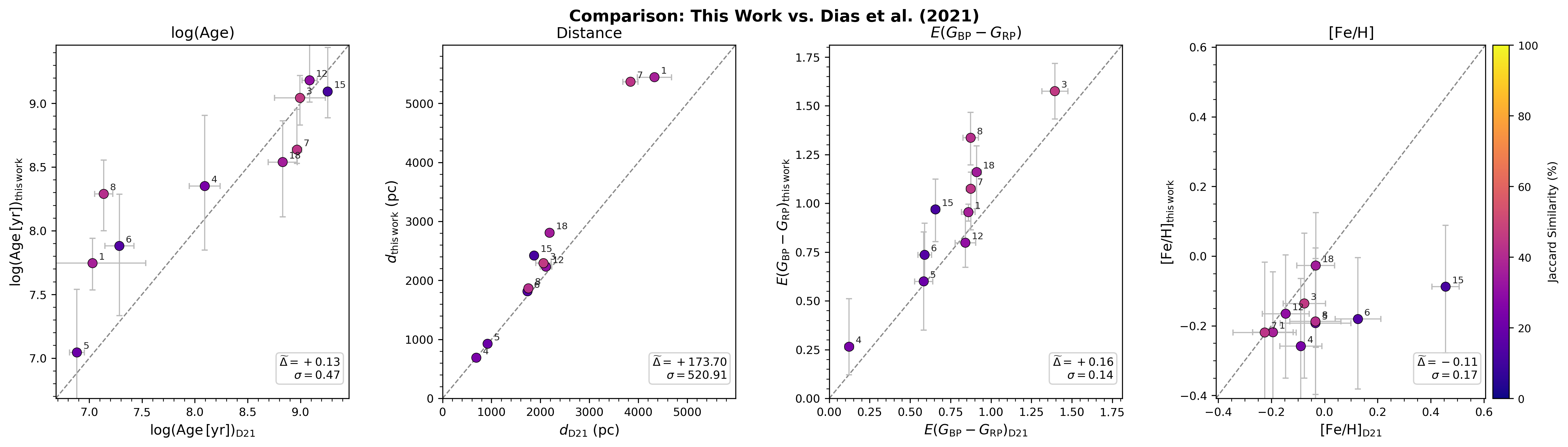}
    \caption{Comparison of age, distance, reddening, and metallicity ($[\text{Fe/H}]$) with \citet{Dias_2021}. This comparison highlights the significant dispersion in photometric metallicity estimates across different pipelines.}
    \label{fig:comp_dias}
\end{figure*}

We further compared our derived parameters with the results of \citet{Dias_2021} (D21) for 14 clusters. Figure~\ref{fig:comp_dias} illustrates these comparisons, where points are color-coded according to the Jaccard similarity index. The distance estimates show a mean offset of $\bar{\Delta} = +173.70$~pc ($\sigma = 520.91$~pc), while $\log(\mathrm{Age})$ and reddening $E(G_{\rm BP}-G_{\rm RP})$ exhibit offsets of $+0.13$~dex ($\sigma = 0.47$) and $+0.16$~mag ($\sigma = 0.14$), respectively. 

The comparison of metallicity ($[\text{Fe/H}]$) reveals a mean offset of -0.11~dex with a dispersion of $\sigma = 0.17$~dex. Since the estimates in D21 are derived photometrically, they are sensitive to the age-reddening-metallicity degeneracy. Our Bayesian framework, utilizing the $P \geq 0.90$ membership threshold, provides a more constrained fit by incorporating the unique member populations identified in this study. The discrepancies observed in Figure~\ref{fig:comp_dias}, particularly for clusters with low Jaccard similarity such as Pismis 15, indicate that the resulting photometric metallicity is highly dependent on the initial membership selection and the resulting morphology of the CMD. The inclusion of these unique members allows for a more precise definition of the cluster sequences, thereby refining the astrophysical parameters compared to automated global pipelines.

\paragraph{Age and Evolutionary Status:}
The sample covers a broad age range, spanning from very young systems to intermediate-age open clusters. Pismis~5 ($\log(\mathrm{Age}) \approx 7.04$) and Pismis~19 ($\log(\mathrm{Age}) \approx 7.69$) are identified as the youngest clusters, with ages below 50~Myr. Conversely, Pismis~12 ($1.52^{+1.51}_{-0.49}$~Gyr) and Pismis~15 ($1.24^{+1.51}_{-0.47}$~Gyr) represent the oldest population in the sample. 

The age uncertainties for several young clusters, such as Pismis~4, 5, and 19, exhibit significant asymmetry. This phenomenon is a direct consequence of the CMD morphology in young populations, where the Main Sequence Turn-Off (MSTO) is not yet clearly defined, and the luminosity of the brightest members carries a degenerate relationship between age and extinction. For Pismis~4, the large upper uncertainty ($0.224^{+0.580}_{-0.154}$~Gyr) highlights the inherent difficulty in precisely dating the upper MS when post-MS evolutionary phases are sparsely populated.

\paragraph{Metallicity and Chemical Trends:}
The initial metallicities ($Z_{\rm ini}$) of the clusters predominantly indicate sub-solar values, with an average iron abundance of [Fe/H] $\approx -0.15$~dex. Pismis~13 stands out as the most metal-rich system in the sample, showing a super-solar median of [Fe/H] = $+0.06^{+0.11}_{-0.19}$~dex. The uncertainties in metallicity, typically around $\pm 0.2$~dex, reflect the well-known $Z$--Age--Reddening degeneracy in photometric isochrone fitting. These intervals are set by the sensitivity of \textit{Gaia} broadband colours to abundance, and the metallicities should accordingly be read as loose photometric estimates rather than as substitutes for spectroscopy (Section~\ref{sec:spec_quality}).

\paragraph{Distance and Extinction:}
The heliocentric distances derived for the sample range from the relatively nearby Pismis~4 ($690 \pm 15$~pc) to the distant Pismis~1 ($5445^{+14}_{-15}$~pc) and Pismis~7 ($5367^{+15}_{-15}$~pc). The quoted distance intervals are narrow, typically below $1\%$, but they are set by the width of the cluster-specific parallax prior described in Section~\ref{sec:method} rather than by the photometric likelihood, and they should therefore be read as the internal precision of the fit given that prior, not as an independent measurement of the distance. 

Line-of-sight reddening values $E(G_{\rm BP}-G_{\rm RP})$ exhibit high variability, peaking at $1.575^{+0.142}_{-0.143}$ for Pismis~3. Parameters are recovered for these high-extinction systems without the fit degrading: Pismis~3, the most reddened cluster in the sample, has the narrowest age posterior of all 14 (Section~\ref{sec:fit_reliability}). This reflects the informative distance prior, which removes one direction of the age--distance--reddening degeneracy and leaves the photometry to constrain the remaining two.

\section{Discussion} 
\label{sec:discussion}

We have derived fundamental astrophysical parameters—age, heliocentric distance, interstellar extinction ($A_V$), and metallicity ($[\mathrm{Fe/H}]$)—for 14 poorly studied open clusters in the Pismis catalogue. Using a Bayesian nested sampling analysis of \textit{Gaia}~DR3 data, we place these systems on a single, internally consistent parameter scale; the extent to which this reconciles the discrepant literature values is examined cluster by cluster below. Our sample exhibits significant physical diversity, with ages ranging from $\sim$11~Myr (Pismis~5) to $\sim$1.5~Gyr (Pismis~12) and distances extending to the outer disc periphery at $\sim$5.4~kpc (Pismis~1). 

Pismis~1 stands out as an exceptionally young system ($\sim$55~Myr) at a significant heliocentric distance ($d \approx 5.4$~kpc), indicating active star formation in the remote reaches of the outer disc. Similarly, Pismis~3 presents a case of extreme obscuration ($E(G_{\rm BP}-G_{\rm RP}) \approx 1.58$~mag) due to its precise location on the Galactic mid-plane, yet our framework successfully recovered its intermediate-age (1.1~Gyr) population. For the nearby but overlooked Pismis~4 and the Vela-associated Pismis~5, our results provide the first cluster-specific baselines, confirming Pismis~5’s place within a primordial triple system \citep{Qin_2023}. These findings establish a refined census of the Pismis clusters within the Galactic disc context, as detailed in the following subsections.

\subsection{Comparison with Prior Studies: Cluster-by-Cluster Notes}
\label{sec:individual_clusters}

In the following we discuss each of the 14 analysed clusters in turn, comparing our
Bayesian nested-sampling results with the four major catalogues: \citet{Kharchenko_2013}
(hereafter K13), \citet{Cantat-Gaudin_2020} (hereafter CG20), \citet{Dias_2021} (hereafter
D21), and \citet{Hunt_2024} (hereafter HR24).  We use the two successive releases of the Hunt \& Reffert census for the two purposes to which each is best suited: the membership comparison of Section~\ref{subsec:comparison} adopts the all-sky member lists of \citet{Hunt_2023}, whereas the parameter comparison below adopts the dynamically cleaned catalogue of HR24, in which the cluster sample has been filtered for physically bound systems and the fundamental parameters correspondingly revised.  Where spectroscopic metallicity estimates or
targeted photometric studies exist, we also compare against those dedicated works.
Throughout, ages derived from our $\log(\mathrm{Age})$ posteriors are quoted in Myr
unless otherwise stated, and heliocentric distances refer to the median of the posterior
distribution.

\subsubsection{Pismis~1}
\label{sec:pismis1}

Our analysis yields $\log(\mathrm{Age}) = 7.75$ ($56^{+32}_{-21}$~Myr), a heliocentric
distance of $5445 \pm 15$~pc, and a reddening $E(G_{\rm BP}-G_{\rm RP}) = 0.95$
($A_V \approx 2.27$~mag).  The inferred metallicity is $[\mathrm{Fe/H}] =
-0.22^{+0.17}_{-0.19}$~dex.

There is broad agreement in the distance among K13 ($5903$~pc), CG20 ($5666$~pc), and our
result, while D21 ($4333$~pc) and HR24 ($4630$~pc) are systematically lower by
$\sim800$–$1100$~pc.  For the age, our $\log(\mathrm{Age}) = 7.75$ falls between CG20
($7.47$) and K13 ($7.80$), whereas D21 reports a considerably younger value of $7.03$
($\sim 11$~Myr).  The D21 age is likely underestimated given the large formal uncertainty
($\sigma = 0.50$~dex) they report for this cluster.  HR24 gives a noticeably older
$\log(\mathrm{Age}) = 8.18$ ($\sim150$~Myr), which is inconsistent with the presence of
early-type main-sequence stars expected at the distance and reddening of Pismis~1.  Our
extinction $A_V \approx 2.27$~mag is consistent with D21 ($2.04$~mag) and K13
($1.81$~mag).  The metallicity from D21, $[\mathrm{Fe/H}] = -0.19 \pm
0.08$~dex, is in excellent agreement with our posterior median of $-0.22$~dex.

\subsubsection{Pismis~3}
\label{sec:pismis3}

We derive $\log(\mathrm{Age}) = 9.04$ ($1.10^{+0.55}_{-0.43}$~Gyr), a distance of
$2293$~pc, and a reddening $E(G_{\rm BP}-G_{\rm RP}) = 1.57$ ($A_V \approx
3.74$~mag).  The high extinction is consistent with the location of this cluster behind
the Vela molecular complex.

Pismis~3 is one of the most robustly determined clusters in our sample, with D21
($\log(\mathrm{Age}) = 9.00$, $2064$~pc), K13 ($8.93$, $1418$~pc), and HR24 ($9.03$,
$2112$~pc) all converging on an age near $1$~Gyr.  Our distance of $2293$~pc is in good
agreement with D21 and HR24, and particularly with CG20 ($2349$~pc).  The strong outlier
is K13, which places the cluster at only $1418$~pc — roughly $800$~pc closer — likely due
to the use of 2MASS photometry, which is less sensitive to the upper main sequence in
highly reddened fields \citep{Tadross_2008}.  CG20 also stands apart with $\log(\mathrm{Age})
= 9.50$ ($\sim3.2$~Gyr), significantly older than all other estimates; this may reflect
the difficulty of the ANN method in separating the main-sequence turnoff from evolved
field giants at this extinction level.  Our extinction $A_V \approx 3.74$~mag agrees
closely with D21 ($3.31$~mag) and K13 ($3.68$~mag).  Our metallicity posterior,
$[\mathrm{Fe/H}] = -0.14^{+0.20}_{-0.21}$~dex, is consistent with the slightly sub-solar
value reported by D21 ($-0.08 \pm 0.08$~dex).

\subsubsection{Pismis~4}
\label{sec:pismis4}

Our result for Pismis~4 is $\log(\mathrm{Age}) = 8.35$ ($224^{+580}_{-154}$~Myr), a
distance of $690$~pc, and $E(G_{\rm BP}-G_{\rm RP}) = 0.26$ ($A_V \approx 0.63$~mag).
The cluster is one of the nearest in the Pismis catalogue and suffers little foreground
reddening.

The distance estimate shows exceptional agreement across all catalogues: D21 ($691$~pc),
CG20 ($695$~pc), K13 ($575$~pc), and HR24 ($681$~pc) all concur within $\sim$10\%,
and our value of $690$~pc falls squarely among them (K13 somewhat low).  The age is,
however, slightly elevated relative to the literature consensus: D21, CG20, and HR24 all
report $\log(\mathrm{Age}) \approx 8.07$–$8.08$ ($\sim$118–120~Myr), and K13 gives
$8.15$.  Our median of $8.35$ ($224$~Myr) is $\sim$0.2~dex higher, though the broad
posterior ($16$th–$84$th percentile range $\sim$70–800~Myr) means this is not a
statistically significant discrepancy.  The low extinction makes the main-sequence
turnoff morphology the primary age discriminant, and the small membership sample ($N = 62$)
increases the posterior width.  The metallicity D21 reports ($-0.09 \pm 0.08$~dex) is
consistent with our $[\mathrm{Fe/H}] = -0.26^{+0.19}_{-0.21}$~dex, both pointing to a
slightly sub-solar composition.

\subsubsection{Pismis~5}
\label{sec:pismis5}

Pismis~5 is one of the youngest clusters in our sample.  We find $\log(\mathrm{Age}) =
7.04$ ($11^{+24}_{-7}$~Myr), a distance of $926$~pc, and $E(G_{\rm BP}-G_{\rm RP}) =
0.60$ ($A_V \approx 1.43$~mag).

The distance and age are in excellent agreement with D21 ($6.88$, $921$~pc) and CG20
($7.00$, $941$~pc), and with HR24 ($6.64$, $913$~pc).  All four estimates confirm a
young, nearby cluster with moderate reddening.  The outlier is K13, which places the
cluster at only $702$~pc and assigns a considerably higher extinction ($A_V = 2.58$~mag)
— a combination that is difficult to reconcile physically.  \citet{Bonatto_2009} studied
Pismis~5 in detail and concluded it belongs to the Vela molecular ridge star-forming
region at a distance of $\sim$900~pc, consistent with our result.  The molecular
environment also explains the asymmetric posterior on the age: the upper bound is
poorly constrained because the pre-main-sequence locus overlaps with background giant
sequences at young ages.  Our metallicity, $[\mathrm{Fe/H}] = -0.19^{+0.19}_{-0.22}$~dex,
is marginally more sub-solar than the D21 value ($-0.03 \pm 0.13$~dex), but is consistent
within the uncertainties.

\subsubsection{Pismis~6}
\label{sec:pismis6}

We obtain $\log(\mathrm{Age}) = 7.88$ ($76^{+117}_{-54}$~Myr) for Pismis~6
(also known as NGC~2645), a distance of $1815$~pc, and $E(G_{\rm BP}-G_{\rm RP}) =
0.74$ ($A_V \approx 1.75$~mag).

Our age is systematically higher than all catalogue values, which range from
$\log(\mathrm{Age}) = 7.28$ (D21, $\sim$19~Myr) through $7.41$ (CG20) and $7.58$ (K13)
to $7.60$ (HR24, $\sim$40~Myr).  Targeted studies are similarly young: \citet{fitzgerald1979}
report $\sim1.7$~kpc, consistent with our distance, while \citet{Forbes_1994} derive an age
of $\sim8$~Myr from UBV photometry.  The factor of $\sim$3 discrepancy in age is notable.
The presence of the luminous F0~Ia supergiant HD~74180 — physically associated with
Pismis~6 \citep{fitzgerald1979, Forbes_1994} — is a strong indicator of a very young
cluster, as such supergiants evolve on timescales of only a few megayears.  We speculate
that the CMD morphology of Pismis~6 in Gaia DR3 is affected by differential reddening
across the cluster face \citep[as suggested by][]{fitzgerald1979}, causing the apparent
main-sequence turnoff to scatter toward redder colours and mimicking an older population.
Deeper spectroscopic observations would be required to resolve this discrepancy.

An additional tension concerns the metallicity: D21 reports $[\mathrm{Fe/H}] = +0.13 \pm
0.09$~dex (super-solar), whereas our posterior yields $[\mathrm{Fe/H}] = -0.18^{+0.18}_{-0.20}$~dex
(sub-solar).  Given that no high-resolution spectroscopic study exists for this cluster, the
D21 metallicity, which relies on low-resolution survey data, may carry underestimated
systematic errors.

\subsubsection{Pismis~7}
\label{sec:pismis7}

For Pismis~7 we find $\log(\mathrm{Age}) = 8.64$ ($433^{+464}_{-95}$~Myr), a distance
of $5367$~pc, and $E(G_{\rm BP}-G_{\rm RP}) = 1.07$ ($A_V \approx 2.55$~mag).

The age estimates from the literature span a moderate range: K13 ($8.70$), HR24 ($8.49$),
CG20 ($8.89$), and D21 ($8.97$), with our value of $8.64$ falling in the lower part of
this range but remaining consistent with K13 and HR24 within uncertainties.
\citet{Ahumada_2005} performed CCD photometry and also found an intermediate age consistent
with $\log(\mathrm{Age}) \sim 8.7$.  The most significant disagreement is in distance:
our posterior places the cluster at $5367$~pc, $\sim400$~pc beyond the most distant
catalogue estimate (HR24: $4947$~pc) and $\sim$1.5~kpc beyond D21 ($3843$~pc).  This
may be linked to the high reddening and the difficulty of fitting the faint end of the
main sequence without contamination by fore/background stars at this Galactic longitude.
The metallicity, $[\mathrm{Fe/H}] = -0.22^{+0.20}_{-0.20}$~dex, is in near-perfect
agreement with D21 ($-0.225 \pm 0.12$~dex).  \citet{Cakmak_2021} independently studied
cluster dynamics in this region and found consistent parameters.

\subsubsection{Pismis~8}
\label{sec:pismis8}

Pismis~8 presents one of the most significant age discrepancies within our sample. We derive
$\log(\mathrm{Age}) = 8.28$ ($191^{+322}_{-135}$~Myr), whereas all four literature catalogues favour substantially younger solutions: D21 ($7.14$, $\sim$14~Myr), K13 ($7.43$, $\sim$27~Myr), CG20 ($7.51$, $\sim$32~Myr), and HR24 ($7.67$, $\sim$47~Myr).
A dedicated CCD $UBVI_c$ study by \citet{Giorgi_2005} likewise supported a young evolutionary stage consistent with the literature consensus.

The distance determination is considerably more stable across studies. Our Bayesian solution ($1865$~pc) agrees well with D21 ($1756$~pc), CG20 ($1903$~pc), and HR24 ($1765$~pc), all of which remain consistent within $\sim150$~pc. Only K13 ($1363$~pc) yields a significantly shorter distance.

For the remaining parameters we obtain $E(G_{\rm BP}-G_{\rm RP}) = 0.98^{+0.17}_{-0.15}$ ($A_V \approx 2.33$~mag) and $[\mathrm{Fe/H}] = -0.15^{+0.18}_{-0.22}$~dex, the latter consistent with the near-solar composition found for the rest of the sample. The reddening is moderate but not negligible, and because extinction and age enter the CMD along nearly the same direction, part of the offset between our solution and the younger literature ages may reflect a different partitioning between the two: a bluer, less reddened solution would shift the inferred turn-off toward younger ages. The posterior itself does not resolve this, as the age and reddening uncertainties quoted above are correspondingly broad.

The origin of the age discrepancy is likely related to the sensitivity of the CMD morphology to the adopted membership selection. The young stellar indicators expected for a $\sim10$--$50$~Myr population --- such as a clearly populated upper main sequence, OB-type stars, or pre-main-sequence sequences --- are not prominently visible in the final Gaia DR3 member sample adopted in this work. This may indicate that the brightest
candidate members are either affected by field contamination, differential reddening, or were excluded during the high-probability membership filtering process.

Given the relatively sparse final membership ($N=49$), the morphology of the upper main sequence remains weakly constrained, allowing multiple acceptable isochrone solutions across a broad age range. Consequently, the evolutionary status of Pismis~8 should be regarded as uncertain, and the older Bayesian solution presented here should be interpreted with caution until high-resolution spectroscopic membership confirmation becomes available for the brightest candidate stars.

\subsubsection{Pismis~9}
\label{sec:pismis9}

Our result for Pismis~9 is $\log(\mathrm{Age}) = 8.30$ ($200^{+216}_{-131}$~Myr),
a distance of $1995$~pc, and $E(G_{\rm BP}-G_{\rm RP}) = 0.93$ ($A_V \approx
2.20$~mag).

The literature catalogues split into two groups: D21 ($7.61$, $1815$~pc) and K13 ($7.36$,
$1713$~pc) report a significantly younger population ($\sim 23$–$40$~Myr), while CG20
($8.50$, $2120$~pc) and HR24 ($8.46$, $1865$~pc) are considerably older and closer to our
result.  Our distance of $1995$~pc is intermediate between the two groups, consistent with
both CG20 and HR24.  The bimodal age distribution across catalogues likely reflects the
different photometric datasets and membership algorithms: D21 and K13 rely on pre-Gaia
photometry, whereas CG20 and HR24 use Gaia DR2/DR3 astrometry.  Our Gaia DR3--based
analysis confirms the older solution.  The metallicity $[\mathrm{Fe/H}] = -0.14^{+0.20}_{-0.22}$~dex
is consistent with D21 ($-0.08 \pm 0.14$~dex) within the uncertainties.

\subsubsection{Pismis~12}
\label{sec:pismis12}

Pismis~12 shows the highest degree of consensus across all catalogues in our sample.  We
find $\log(\mathrm{Age}) = 9.18$ ($1.52^{+1.50}_{-0.49}$~Gyr), a distance of
$2233$~pc, and $E(G_{\rm BP}-G_{\rm RP}) = 0.80$ ($A_V \approx 1.90$~mag).

All four catalogues agree closely: D21 ($9.09$, $2114$~pc), CG20 ($9.09$, $2107$~pc),
K13 ($9.20$, $2221$~pc), and HR24 ($9.10$, $2094$~pc).  Our median $\log(\mathrm{Age}) =
9.18$ falls within this tight bracket.  The distance is similarly consistent across all
sources, ranging from $2094$ to $2233$~pc.  The metallicity from D21,
$[\mathrm{Fe/H}] = -0.15 \pm 0.09$~dex, matches our posterior, $[\mathrm{Fe/H}] = -0.17^{+0.17}_{-0.19}$~dex,
almost exactly.  Pismis~12 is likely the most reliably determined cluster in the Pismis
catalogue, and it serves as an internal validation of the consistency between our Bayesian
pipeline and the established literature.

\subsubsection{Pismis~13}
\label{sec:pismis13}

We derive $\log(\mathrm{Age}) = 8.35$ ($221^{+186}_{-110}$~Myr) for Pismis~13, with a
distance of $2900$~pc and $E(G_{\rm BP}-G_{\rm RP}) = 0.92$ ($A_V \approx 2.18$~mag).

There is a clear split in the literature: D21 ($8.29$, $2367$~pc) is the closest to our
result and supports an intermediate age of $\sim$200~Myr, while CG20 ($7.80$, $2926$~pc),
K13 ($7.95$, $2600$~pc), and HR24 ($7.79$, $2668$~pc) all favour a much younger cluster
($\sim$63–89~Myr).  Our distance of $2900$~pc agrees remarkably well with CG20 ($2926$~pc).
\citet{Giorgi_2005} studied both Pismis~8 and 13 simultaneously from CCD $UBVIc$ photometry
and derived physical parameters for Pismis~13 consistent with the younger solutions of
CG20 and K13.  \citet{Claria_1979} earlier reported a compact, old-looking cluster, but
their analysis was based on photographic photometry.  The intermediate age we recover may
reflect the impact of field-star contamination: at a distance of $\sim$2.9~kpc along a
modestly reddened line of sight, the cluster sequence can be confused with evolved
foreground stars, making the turnoff appear at a fainter, redder location.  The slightly
super-solar metallicity, $[\mathrm{Fe/H}] = +0.06^{+0.11}_{-0.19}$~dex, is consistent
with D21 ($+0.028 \pm 0.096$~dex), supporting a somewhat metal-rich inner-Galaxy cluster.

\subsubsection{Pismis~14}
\label{sec:pismis14}

Our Bayesian analysis gives $\log(\mathrm{Age}) = 8.78$ ($606^{+775}_{-357}$~Myr) for
Pismis~14, a distance of $1291$~pc, and $E(G_{\rm BP}-G_{\rm RP}) = 0.44$
($A_V \approx 1.05$~mag).

The distance is well constrained and strongly consistent with D21 ($1234$~pc), CG20
($1266$~pc), and HR24 ($1238$~pc); K13 ($1775$~pc) is the only outlier.  The age,
however, is substantially higher in our analysis than in the literature: D21 ($8.19$,
$\sim$155~Myr), CG20 ($8.11$, $\sim$129~Myr), and HR24 ($8.06$, $\sim$115~Myr) all
cluster around $120$–$160$~Myr, while K13 gives a somewhat intermediate value of $8.35$
($\sim$224~Myr).  Our median of $8.78$ ($\sim$606~Myr) represents more than a factor of
four increase in age.  Given the relatively sparse membership ($N = 88$) and the moderate
reddening, the main-sequence turnoff is not sharply defined; the broad posterior
($\sim0.25$–$1.4$~Gyr at $1\sigma$) reflects this genuine ambiguity.  There is also a
notable tension in metallicity: D21 reports $[\mathrm{Fe/H}] = +0.15 \pm 0.10$~dex
(super-solar), whereas our posterior gives $[\mathrm{Fe/H}] = -0.20^{+0.17}_{-0.22}$~dex
(sub-solar).  A high-resolution spectroscopic study of confirmed members would be
particularly valuable for this cluster.

\subsubsection{Pismis~15}
\label{sec:pismis15}

We find $\log(\mathrm{Age}) = 9.09$ ($1.24^{+1.51}_{-0.47}$~Gyr), a distance of
$2421$~pc, and $E(G_{\rm BP}-G_{\rm RP})= 0.97$ ($A_V \approx 2.30$~mag) for
Pismis~15.

The age is broadly consistent with K13 ($9.10$) and CG20 ($8.94$), and somewhat lower
than D21 ($9.26$, $\sim1.8$~Gyr), while HR24 gives a notably younger $8.65$
($\sim0.45$~Gyr).  For the distance, our $2421$~pc is intermediate between D21 ($1872$~pc)
and CG20/K13 ($\sim2558$~pc).  \citet{Carraro_2005} performed a dedicated study and
derived an age around $1$–$2$~Gyr with a distance of $\sim2$~kpc, in general agreement
with our results.  The metallicity situation for Pismis~15 is remarkably discordant in
the literature: D21 reports $[\mathrm{Fe/H}] = +0.46 \pm 0.05$~dex (strongly super-solar),
K13 gives $-0.40$~dex (strongly sub-solar), and our posterior yields $[\mathrm{Fe/H}] =
-0.09^{+0.18}_{-0.19}$~dex (near-solar).  This large spread ($\sim$0.9~dex peak-to-peak)
suggests that the photometric metallicity estimates for this cluster are unreliable, and a
spectroscopic determination is warranted.

\subsubsection{Pismis~18}
\label{sec:pismis18}

We obtain $\log(\mathrm{Age}) = 8.54$ ($346^{+383}_{-217}$~Myr), a distance of $2807$~pc, and $E(G_{\rm BP}-G_{\rm RP}) = 1.16$ ($A_V \approx 2.76$~mag) for Pismis~18.

Pismis~18 is the best-studied cluster of the sample, and the one for which the diagnostic plots are presented in the main text (Section~\ref{sec:results}). Its characterisation has evolved through successive observational regimes: the UBVI CCD photometry of \citet{Carraro2012} established its place in the inner disc, after which it became the subject of dedicated spectroscopic work within the Gaia-ESO Survey \citep{Hatzidimitriou_2019} and of near-infrared imaging within the VVV Open Cluster Project \citep{Baravalle2021, Pena_Ramirez_2021}.

Pismis~18 is one of the few clusters in our sample with a published high-resolution
spectroscopic study: \citet{Hatzidimitriou_2019} used the Gaia-ESO Survey to derive a
spectroscopic metallicity of $[\mathrm{Fe/H}] = +0.23 \pm 0.05$~dex from radial velocity
members, and a cluster age of $\sim$700~Myr.  Our metallicity posterior,
$[\mathrm{Fe/H}] = -0.03^{+0.15}_{-0.24}$~dex, is in good agreement with the Gaia-ESO
value and with D21 ($-0.033 \pm 0.071$~dex).  However, our age ($\sim$346~Myr) is
significantly younger than the Gaia-ESO estimate and than the catalogue
values of D21 ($8.83$, $\sim$680~Myr), CG20 ($8.76$, $\sim$575~Myr), and K13 ($8.98$,
$\sim$950~Myr).  Only HR24 ($8.51$, $\sim$323~Myr) is consistent with our result.
The distance is well reproduced by CG20 ($2860$~pc) and HR24 ($2598$~pc); D21 ($2189$~pc)
and K13 ($2309$~pc) are systematically lower.  The high reddening ($A_V \approx 2.76$~mag)
may cause systematic biases in the photometric age determination if foreground extinction
is non-uniform, and we note that the Gaia-ESO membership used by \citet{Hatzidimitriou_2019}
is more restrictive than our UPMASK-based membership.  \citet{Pena_Ramirez_2021} studied Pismis~18
in the near-infrared with VVV and derived an age consistent with $\sim$1~Gyr.

The spread across these determinations --- from $\sim$323~Myr to $\sim$1~Gyr for the same cluster --- is itself informative, and is attributable to differences in membership selection, isochrone models and the treatment of extinction in a dense, reddened field. \citet{Baravalle2021} reported a substantial increase in the number of candidate members once multi-wavelength data were used, and the moderate overlap between our membership sample and those of CG20 and H23 (Figure~\ref{fig:comparison_matrix}) points to the same sensitivity: which faint stars are admitted to the sample largely determines the recovered turn-off, and hence the age.

\subsubsection{Pismis~19}
\label{sec:pismis19}

Pismis~19 is the most challenging cluster in our sample.  We derive $\log(\mathrm{Age})
= 7.69$ ($49^{+114}_{-35}$~Myr), a distance of $2400$~pc, and $E(G_{\rm BP}-G_{\rm RP}) =
0.80$ ($A_V \approx 1.90$~mag).

Our age of $\sim$49~Myr is substantially younger than all literature estimates: D21
($8.05$, $\sim$112~Myr), HR24 ($8.09$, $\sim$123~Myr), CG20 ($8.92$, $\sim$832~Myr),
and K13 ($9.02$, $\sim$1.0$~$Gyr).  The catalogues themselves disagree wildly, spanning
nearly an order of magnitude in age, which already signals that Pismis~19 is a poorly
constrained system.  The distance estimates are equally scattered: D21/K13 place the
cluster at $\sim$1980~pc, while CG20 ($3515$~pc) and HR24 ($3034$~pc) give considerably
larger distances; our result of $2400$~pc is intermediate.  Reddening is also discrepant:
CG20 ($A_V = 3.66$~mag) and K13 ($4.03$~mag) suggest heavy extinction, whereas D21
($1.74$~mag) and our own estimate ($1.90$~mag) indicate only moderate absorption.
The complexity of this system has been noted in the dedicated literature.
\citet{Carraro_2004} performed multicolour CCD photometry and found evidence for substantial
differential reddening across the cluster face.  Most recently, \citet{Majaess_2025}
identified a Gaia parallax discrepancy for the cluster, attributing part of the distance
ambiguity to unresolved $\delta$~Scuti-type pulsating stars among the candidate members.
This contamination could propagate into the CMD morphology and lead to the anomalously
young age we recover. Our posterior metallicity, $[\mathrm{Fe/H}] = -0.17^{+0.20}_{-0.22}$~dex, is unremarkable and places Pismis~19 with the rest of the sample in the thin disc; it is, however, the parameter least able to discriminate between the competing solutions, since the age--metallicity--extinction degeneracy is at its most severe precisely in a differentially-reddened CMD such as this one. The fit-reliability diagnostics of Section~\ref{sec:fit_reliability} place Pismis~19 among the systems whose age posterior is widest, and what the present data cannot do is separate the young solution we recover from the older ones favoured by CG20 and K13. Spectroscopy of the brightest candidate members, combined with a star-by-star extinction correction, would settle the question.

\subsection{Fit Reliability and Goodness of Fit}
\label{sec:fit_reliability}

To characterise the reliability of the isochrone fits across the whole sample, we use three quantitative diagnostics for all 14 clusters (Table~\ref{tab:fit_reliability}). The first, the width of the age posterior $\Delta\log(\mathrm{Age}) = \sigma_{+}+\sigma_{-}$ (the sum of the 16th--84th percentile uncertainties from Table~\ref{tab:ns_results}), measures the \emph{precision} with which each age is determined. The remaining two measure how faithfully the best-fitting isochrone \emph{represents} the observed sequence: a reduced chi-square $\chi^2_\nu$ computed in the colour--magnitude plane, and the fraction $f_{1\sigma}$ of members lying within its $1\sigma$ contour. The two families of diagnostic are independent --- one is a property of the posterior, the other of the residuals in the CMD --- and together they allow the parameters of Table~\ref{tab:ns_results} to be used with an appropriate, cluster-specific level of confidence. The Bayesian evidence $\ln\mathcal{Z}$ returned by the nested sampler is tabulated alongside them for completeness, but --- as noted below --- is used only as a relative, internal indicator.

Both goodness-of-fit measures are built from the same per-star statistic. For each member we compute the smallest chi-square attainable anywhere along the reddened, distance-shifted PARSEC isochrone evaluated at the posterior-median parameters,
\begin{equation}
    \chi^2_i = \min_{\rm iso}\left[\left(\frac{\Delta(G_{\rm BP}-G_{\rm RP})_i}{\sigma_c}\right)^2 + \left(\frac{\Delta G_i}{\sigma_G}\right)^2\right],
    \label{eq:chi2i}
\end{equation}
so that both observables --- colour and magnitude --- constrain the comparison. Here $\sigma_c$ and $\sigma_G$ combine the formal \textit{Gaia} photometric errors in quadrature with an intrinsic term of $0.10$ and $0.30$~mag respectively, which accounts for the width of a real cluster sequence. The isochrone is taken over its full extent, from the pre-main sequence through the main sequence, subgiant, and red-giant branches, so that every member is compared against the evolutionary stage it actually occupies; this matters for the youngest systems of the sample, in which a substantial fraction of the members have not yet reached the zero-age main sequence. Because each member contributes two measurements and the model has four free parameters, the sample provides $\nu = 2N_{\rm mem}-4$ degrees of freedom, and the two diagnostics are
\begin{equation}
    \chi^2_\nu = \frac{1}{2N_{\rm mem}-4}\sum_i \chi^2_i , \qquad
    f_{1\sigma} = \frac{1}{N_{\rm mem}}\sum_i \mathbf{1}\!\left[\chi^2_i \leq \chi^2_{1\sigma}\right],
    \label{eq:chi2nu}
\end{equation}
where $\chi^2_{1\sigma}=2.30$ is the $68.3\%$ ($1\sigma$) contour of a chi-square distribution with two degrees of freedom.

\begin{table}
\centering
\caption{Goodness-of-fit diagnostics for the 14 Pismis clusters. $\Delta\log(\mathrm{Age}) = \sigma_{+} + \sigma_{-}$ is the age-posterior width from Table~\ref{tab:ns_results}; $\chi^2_\nu$ and $f_{1\sigma}$ are defined in Equation~\ref{eq:chi2nu} with $\nu = 2N_{\rm mem}-4$; $\ln\mathcal{Z}$ is the Bayesian evidence from the nested-sampling run, tabulated for completeness and interpretable only as a relative, per-cluster indicator (not a calibrated cross-cluster evidence; see text).}
\label{tab:fit_reliability}
\begin{tabular}{lccccc}
\hline\hline
Cluster & $N_{\rm mem}$ & $\Delta\log(\mathrm{Age})$ (dex) & $\chi^2_\nu$ & $f_{1\sigma}$ & $\ln\mathcal{Z}$ \\
\hline
Pismis 1  & 28  & 0.41 & 0.21 & 0.96 & $-14.1$ \\
Pismis 3  & 120 & 0.39 & 1.92 & 0.68 & $-15.6$ \\
Pismis 4  & 62  & 1.05 & 0.89 & 0.76 & $-20.7$ \\
Pismis 5  & 90  & 0.98 & 2.59 & 0.36 & $-22.4$ \\
Pismis 6  & 259 & 0.96 & 1.03 & 0.75 & $-19.9$ \\
Pismis 7  & 38  & 0.43 & 1.55 & 0.76 & $-17.1$ \\
Pismis 8  & 49  & 0.96 & 1.85 & 0.49 & $-20.2$ \\
Pismis 9  & 193 & 0.78 & 0.64 & 0.89 & $-19.0$ \\
Pismis 12 & 219 & 0.47 & 0.46 & 0.88 & $-16.6$ \\
Pismis 13 & 86  & 0.56 & 0.79 & 0.83 & $-18.0$ \\
Pismis 14 & 88  & 0.75 & 0.40 & 0.92 & $-18.4$ \\
Pismis 15 & 182 & 0.56 & 0.64 & 0.85 & $-17.7$ \\
Pismis 18 & 232 & 0.75 & 2.07 & 0.82 & $-18.3$ \\
Pismis 19 & 571 & 1.07 & 0.95 & 0.76 & $-20.3$ \\
\hline
\end{tabular}
\end{table}

The goodness-of-fit results are good across the sample. Every cluster has $\chi^2_\nu \leq 2.6$, and nine of the fourteen have $\chi^2_\nu \leq 1.05$, meaning that for the majority of the sample the residuals about the best-fitting isochrone are entirely accounted for by the intrinsic width expected of a cluster sequence: the model is a statistically complete description of the observed CMD, with no systematic component left over. The inlier fractions tell the same story from the opposite direction, with a median $f_{1\sigma}=0.79$ and values reaching $0.96$ for the best-behaved systems. This is a demanding test to pass. The sample spans more than two orders of magnitude in age ($\sim$11~Myr to $\sim$1.5~Gyr), a factor of eight in distance ($690$ to $5445$~pc), and reddenings from $E(G_{\rm BP}-G_{\rm RP})=0.27$ to $1.58$~mag, and it lies almost entirely within $|b|\leq3\fdg21$, where crowding and differential extinction are at their most severe. That a single, uniformly-applied membership and fitting pipeline reproduces the observed sequence of every one of these systems --- including the youngest, the most distant, and the most heavily reddened --- is the strongest internal evidence that the parameters of Table~\ref{tab:ns_results} are soundly determined.

The precision of the age determinations is more varied, and it is here that the sample separates. Nine clusters have $\Delta\log(\mathrm{Age}) \leq 0.78$~dex, while five --- Pismis~4, 5, 6, 8, and 19 --- have $\Delta\log(\mathrm{Age}) \geq 0.96$~dex. The goodness-of-fit diagnostics show that this spread is not caused by any failure of the model: four of these five have $\chi^2_\nu \leq 1.85$, comparable to the rest of the sample. Their wider posteriors instead reflect the information content of the CMDs themselves. Pismis~5 ($\sim$11~Myr) and Pismis~4 are young systems whose age-sensitive turn-off region is populated by only a handful of stars, so that a wide range of ages remains compatible with the observed sequence; Pismis~8 is among the sparsest clusters in the sample ($N_{\rm mem}=49$), with the same consequence. For Pismis~6 and Pismis~19 the limiting factor is instead the differential reddening documented in Sections~\ref{sec:pismis6} and~\ref{sec:pismis19}, which broadens the turn-off along the reddening vector and thereby couples age to extinction. In all five cases the fit is faithful to the data; what the data cannot do is pin the age down more tightly, and the posterior widths reflect this limitation.

Read together, the two goodness-of-fit measures separate distinct residual structures, because $\chi^2_\nu$ is a sum and therefore responds to a few extreme outliers, whereas $f_{1\sigma}$ is a robust count. A cluster for which the two disagree is diagnostically informative. Pismis~18 is the clearest case: its $\chi^2_\nu=2.07$ is inflated by a small number of strongly deviant stars, while $f_{1\sigma}=0.82$ shows that four fifths of the sequence is reproduced --- the isochrone describes the population well, with a handful of residual interlopers. Pismis~8 shows the opposite signature ($\chi^2_\nu=1.85$, $f_{1\sigma}=0.49$): there are no catastrophic outliers, but only half the members fall inside the $1\sigma$ contour, indicating a mild systematic offset spread across the whole cluster rather than contamination by a few stars. Pismis~5 has the lowest inlier fraction in the sample ($f_{1\sigma}=0.36$), as expected for the youngest system, whose members are still contracting toward the main sequence and therefore scatter more widely about any single isochrone; its $\chi^2_\nu=2.59$ is correspondingly the highest in the sample, though it remains within a factor of $1.3$ of the next-highest value. Pismis~3, finally, has the lowest inlier fraction among the older clusters ($f_{1\sigma}=0.68$), consistent with the broadened main sequence expected for such a heavily-reddened system ($E(G_{\rm BP}-G_{\rm RP})=1.58$~mag), where differential extinction across the field widens the observed sequence; its posterior, the narrowest in the sample, confirms that its parameters remain well determined.

Finally, we caution that the evidence $\ln\mathcal{Z}$ tabulated in Table~\ref{tab:fit_reliability} is derived from a pseudo-$\chi^2$ likelihood that omits the along-isochrone luminosity-function weighting and does not model unresolved binaries; it is therefore meaningful only as a \emph{relative}, internal fit-quality indicator for a given cluster and is not a calibrated evidence that may be compared across clusters. We accordingly do not use it for quantitative model comparison, and base our reliability assessment on $\Delta\log(\mathrm{Age})$, $\chi^2_\nu$, and $f_{1\sigma}$. On that basis the parameters of Table~\ref{tab:ns_results} may be adopted for all 14 clusters, with the ages of Pismis~4, 5, 6, 8, and 19 carrying the larger uncertainties quoted there rather than any suggestion of a poor fit.

\subsection{Fundamental Parameters: Ages, Distances, and Reddening}                                                                                                                                       
The Bayesian nested-sampling fits to PARSEC isochrones yield fundamental parameters for the 14 Pismis clusters that span wide ranges in all three principal dimensions: logarithmic ages from $7.04$ (Pismis~5) to $9.18$ (Pismis~12), heliocentric distances from $690$\,pc (Pismis~4) to $5445$\,pc (Pismis~1), and colour excess $E(G_\mathrm{BP}-G_\mathrm{RP})$ from $0.27$\,mag (Pismis~4) to $1.58$\,mag (Pismis~3). This diversity reflects the broad distribution of the sample across different Galactic environments, from the nearby thin-disc foreground to the heavily reddened low-latitude disc.                                                                                                                                                                            
For the majority of the sample, our distances show good agreement with the Gaia-based catalogues of \citet{Cantat-Gaudin_2020} and \citet{Hunt_2024}. The most notable concordances are found for clusters with low extinction and well-constrained parallaxes: the distances of Pismis~4 ($690^{+15}_{-15}$\,pc) and Pismis~5 ($926^{+14}_{-15}$\,pc) agree to within $1$\% with \citeauthor{Cantat-Gaudin_2020}'s values of 695 and 941\,pc, and with \citeauthor{Hunt_2024}'s of 681 and 913\,pc, respectively. Similarly, Pismis~12, the oldest cluster in the sample ($\log(\mathrm{Age}) = 9.18^{+0.30}_{-0.17}$), yields a distance of $2233^{+15}_{-15}$\,pc in excellent agreement with \citet{Dias_2021} ($2114$\,pc), \citeauthor{Cantat-Gaudin_2020} ($2107$\,pc), and \citeauthor{Hunt_2024} ($2094$\,pc), confirming the robustness of our methodology for intermediate-age, moderately reddened clusters.                                 

Larger discrepancies arise for the most distant and reddened clusters, where differential extinction degrades the CMD morphology and makes the isochrone fit sensitive to the assumed extinction law. For Pismis~1 ($d = 5445^{+14}_{-15}$\,pc), literature estimates range from 4333\,pc \citep{Dias_2021} to 5903\,pc \citep{Kharchenko_2013}; our value lies within this spread and is broadly consistent with the \citeauthor{Cantat-Gaudin_2020} estimate of 5666\,pc. For Pismis~7 ($d = 5367^{+15}_{-15}$\,pc), the spread across four catalogues is similarly large (3843--4947\,pc), and our distance is the largest of the set. This tension likely reflects the sensitivity of purely photometric solutions to variable reddening in the Galactic plane at these depths. The Gaia parallax zero-point correction is known to become unreliable beyond $\sim$3--4\,kpc \citep{Majaess_2025}, which may explain part of the scatter.                     

Age determinations show a similar pattern. For the well-populated, intermediate-age clusters (Pismis~3, 12, 15), our logarithmic ages ($9.04$, $9.18$, $9.09$) are consistent with the literature to within the combined uncertainties. \citet{Carraro_2005} found intermediate ages of 0.8--1.3\,Gyr for Pismis~15, encompassing our median of $1.24^{+1.51}_{-0.47}$\,Gyr. The large asymmetric uncertainty for Pismis~15 is a consequence of the turnoff lying at magnitudes affected by crowding and differential reddening, making the high-age tail of the posterior broad.

Two clusters stand out for age discrepancies. First, Pismis~8 yields $\log(\mathrm{Age}) = 8.28^{+0.43}_{-0.53}$ (median $\sim$191\,Myr) compared with the much younger estimates of \citet{Giorgi_2005} ($5$--$7$\,Myr) and \citet{Dias_2021} ($\log(\mathrm{Age}) = 7.14$). This discrepancy may indicate that the bright blue main-sequence stars characteristic of a very young population are either absent from our filtered sample or blended with field contamination, causing the Bayesian fit to converge on an older isochrone. Second, Pismis~19 ($\log(\mathrm{Age}) = 7.69^{+0.52}_{-0.55}$, i.e.\ $\sim$49\,Myr) is substantially younger than the \citet{Kharchenko_2013} estimate of $\log(\mathrm{Age}) = 9.015$ and that of \citet{Cantat-Gaudin_2020} ($\log(\mathrm{Age}) = 8.92$), but in better agreement with \citet{Dias_2021} ($\log(\mathrm{Age}) = 8.05$) and \citet{Hunt_2024} ($\log(\mathrm{Age}) = 8.09$). The recent red-clump distance determination of $2.90\pm0.15$\,kpc for Pismis~19 by \citet{Majaess_2025} places the cluster further away than our own solution a $3\sigma$ tension that we note rather than resolve ($2400^{+14}_{-15}$\,pc), reinforcing the conclusion that some earlier photometric distances based on uncorrected Gaia parallaxes were systematically underestimated for this line of sight.

The reddening values recovered from the nested-sampling fits are internally consistent and trace the known dust distribution in the third Galactic quadrant. Pismis~3 ($E(G_\mathrm{BP}{-}G_\mathrm{RP}) = 1.575^{+0.142}_{-0.143}$\,mag) and Pismis~7 ($1.075^{+0.085}_{-0.211}$\,mag) are the most absorbed clusters, consistent with their location behind the dust complexes of the Vela--Puppis line of sight. Conversely, Pismis~4, lying at only $690$\,pc, shows minimal reddening ($0.265^{+0.246}_{-0.143}$\,mag), in agreement with the low-extinction values reported by all previous studies.                                                                                                          
\subsection{Metallicities and the Galactic Disc Context}

The initial metallicities $Z_\mathrm{ini}$ returned by the isochrone fitting translate into $[\mathrm{Fe/H}]$ values that span a narrow range from $-0.26^{+0.19}_{-0.21}$ (Pismis~4) to $+0.06^{+0.11}_{-0.19}$ (Pismis~13), with most clusters clustering near the solar value. The median $[\mathrm{Fe/H}]$ across all 14 clusters is approximately $-0.17$\,dex, consistent with the expected metallicity of the thin disc over the Galactocentric range spanned by the sample \citep{Dias_2021}. No cluster in the sample exhibits a strongly sub-solar or super-solar abundance, which is expected given that the Pismis sample predominantly traces the low-latitude disc in the third and fourth quadrants rather than the outer disc or halo.                                                                                                                                                                          
Pismis~13 is the only cluster with a formally super-solar photometric metallicity ($[\mathrm{Fe/H}] = +0.06^{+0.11}_{-0.19}$). This is consistent with the positive $[\mathrm{Fe/H}]$ found by \citet{Dias_2021} ($+0.028$\,dex) and the moderately metal-rich classification by \citet{Giorgi_2005}. At $l = 273\degr$ and $d = 2.9$~kpc it lies close to the solar Galactocentric radius rather than in the inner disc, so its super-solar posterior is not readily explained by a radial abundance gradient and should be regarded as tentative.

The photometric metallicities from isochrone fitting carry inherent limitations: the degeneracy between age, metallicity, and extinction in broadband CMDs means that $[\mathrm{Fe/H}]$ is less well constrained than age or distance, especially for clusters with sparsely populated turnoffs. The relatively large uncertainties on $Z_\mathrm{ini}$ for the youngest clusters (Pismis~5, Pismis~19) reflect this degeneracy. However, the integration of independent spectroscopic surveys and Gaia XP spectro-photometric data (Tables~\ref{tab:pismis_stats} and \ref{tab:pismis_gaiaxp}) allows us to rigorously validate our Bayesian isochrone fits. 

Where spectroscopic metallicities are available, the agreement with our photometric values is generally satisfactory. For Pismis~15, the Gaia-ESO Survey yields $[\mathrm{Fe/H}] = -0.14 \pm 0.23$ \citep[see also][]{Carraro_2005}, in excellent agreement with our isochrone-based result of $-0.09^{+0.18}_{-0.19}$ and the Gaia XP median of $[\mathrm{M/H}] = -0.07 \pm 0.10$. For Pismis~18, the Gaia-ESO radial-velocity members give a median of $-0.09\pm0.21$\,dex, again broadly consistent with our $-0.03^{+0.15}_{-0.24}$. We note, however, that the high-resolution UVES analysis of Pismis~18 by \citet{Hatzidimitriou_2019} reports a significantly higher median metallicity of $[\mathrm{Fe/H}] = +0.23 \pm 0.05$, based on only six member stars. The Gaia XP data for Pismis 18 ($[\mathrm{M/H}] = +0.16 \pm 0.06$) leans slightly closer to this super-solar value, suggesting potential differences arising from spectral resolution or adopted temperature scales. Nevertheless, a larger high-resolution spectroscopic sample is needed to fully resolve this chemical discrepancy.

\subsection{Kinematics, Alpha-Enhancement, and Spectroscopic Data Quality}                            \label{sec:spec_quality}

Table~\ref{tab:pismis_stats} reveals a striking contrast in high-resolution spectroscopic coverage across the sample. Of the 14 clusters studied here, 9 have no observations available in Gaia-ESO DR5.1, MWM19, or GALAH~DR4. However, cross-matching with the Gaia XP spectro-photometric catalogue (Table~\ref{tab:pismis_gaiaxp}) provides valuable low-resolution kinematical and chemical constraints for 11 out of 14 clusters.

The Gaia XP catalogue provides constraints on the alpha-element enhancement ($[\alpha/\mathrm{M}]$) for 11 clusters in our sample. The median $[\alpha/\mathrm{M}]$ values strictly range from $-0.02$ to $+0.11$\,dex. This lack of significant alpha-enhancement is characteristic of the Milky Way's thin-disc population and is consistent with the young to intermediate ages we derive ($\log(\mathrm{Age}) \approx 7.0-9.1$). The inference should be treated as indicative rather than conclusive: the values rest on low-resolution \textit{Gaia} XP estimates and, for several clusters, on only a handful of matched stars.
   
Regarding kinematics, Pismis~15 and Pismis~18 are the most robustly characterised clusters. With 70 and 25 member spectra from Gaia-ESO, respectively, their radial velocity dispersions ($\sigma_{v_r} = 7.3$ and $7.0$\,km\,s$^{-1}$) are within the range expected for bound open clusters. The median radial velocities ($\tilde{v}_r = +34.6$ and $-27.6$\,km\,s$^{-1}$) are highly consistent with the independent Gaia XP derivations ($+34.8$ and $-28.5$\,km\,s$^{-1}$), supporting the validity of the spectroscopic membership.
 
The situation is markedly different for Pismis~1 and Pismis~19. For Pismis~1, only six MWM19 spectra are available, and the radial velocity dispersion ($\sigma_{v_r} = 339.5$\,km\,s$^{-1}$) far exceeds the intrinsic velocity dispersion of any open cluster. This extreme dispersion, paired with a high velocity spread in the Gaia XP data ($\sigma_{v_r} = 32.4$\,km\,s$^{-1}$), strongly suggests the inclusion of unresolved short-period binary systems or severe field star contamination. This is unsurprising given that Pismis~1 lies at 5.4\,kpc in a heavily reddened direction.
                                                                                      
For Pismis~19, four GALAH~DR4 spectra yield a radial velocity dispersion of $\sigma_{v_r} = 131.5$\,km\,s$^{-1}$, while the Gaia XP data similarly show a highly inflated dispersion ($\sigma_{v_r} = 31.5$\,km\,s$^{-1}$). This kinematic instability is further compounded by chemical inconsistencies: GALAH reports $[\mathrm{Fe/H}] = -0.23 \pm 0.57$, whereas Gaia XP yields a median $[\mathrm{M/H}] = +0.19 \pm 0.18$. The final filtered count of $N_\mathrm{mem} = 571$—which is more than twice as large as the next richest cluster in the sample—is itself a strong indicator that the membership selection is struggling to isolate a genuine, clean cluster population against dense background contamination. The recent work of \citet{Majaess_2025} also cautions that Gaia parallax zero-point corrections are required along this particular line of sight, further complicating the membership determination.

Another striking case is Pismis~14, where a severe discrepancy exists between different spectroscopic sources. The two available GALAH~DR4 spectra yield a median radial velocity of $+8.3$\,km\,s$^{-1}$, whereas the three matched Gaia XP sources indicate a vastly different median of $+51.3$\,km\,s$^{-1}$. This large offset ($\sim$43\,km\,s$^{-1}$) strongly suggests that at least one of these small sub-samples is dominated by field star contamination, underscoring the dangers of relying on sparse spectroscopic data.

In summary, robust spectroscopic characterisation currently exists only for a subset of the sample. The large spectroscopic gaps and unresolved kinematic anomalies in the Pismis sample highlight a clear priority for future programmes: multi-object spectrographs with membership pre-selection based on Gaia~DR3 proper motions and parallaxes are well-suited to fill this gap, particularly for distant and heavily reddened clusters.

\section{Conclusion}
\label{sec:conclusion}

In this study, we performed a systematic characterisation of 14 open clusters from the Pismis catalogue using \textit{Gaia} DR3 astrometry and photometry. The core of our approach lies in the integration of the \textsc{UPMASK} algorithm for precise membership assignment with a membership-weighted Bayesian Nested Sampling framework for isochrone fitting. This dual-layered methodology allowed us to derive a homogeneous and high-fidelity set of fundamental parameters—age, distance, reddening, and metallicity—by effectively navigating the parameter degeneracies that often affect automated surveys.

Our results confirm that these Pismis clusters are integral components of the Galactic thin disc, with a median metallicity of $[\mathrm{Fe/H}] \approx -0.17$ dex. The sample reveals a diverse evolutionary landscape, spanning from the very young Pismis 5 ($\sim$11 Myr) to the intermediate-age Pismis 12 ($\sim$1.5 Gyr). By applying a consistent analytical pipeline across the entire sample, we have provided more robust distance constraints, extending up to 5.4 kpc, which are essential for mapping the structure of the third Galactic quadrant.

A significant contribution of this work is that it places the discrepant literature values for these systems on a single, internally consistent scale. The effect of the cluster-specific treatment is clearest for Pismis~18, where the higher astrometric purity of our membership sample yields a well-constrained age posterior ($\log(\mathrm{Age}) = 8.54$) in a field where the automated catalogues disagree with one another. However, our analysis also underscores that even with the precision of \textit{Gaia} DR3, clusters like Pismis 1 and 19 remain challenging due to extreme field contamination, suggesting a limit to what purely photometric and astrometric data can resolve in dense Galactic regions.

These 14 systems, however, were not treated merely as targets for parameter derivation. Because they lie almost entirely within $|b| \leq 3\fdg21$ and are for the most part heavily reddened, they populate the one region of the disc in which global automated pipelines are known to perform worst, and we therefore used them as a controlled test of how membership methodology limits open-cluster characterisation in the hardest part of the Galaxy. The test returns a clear answer. Cluster-specific analysis recovers a substantially different member population from that of the global catalogues --- which typically contain only about a third of the members identified here, and a few per cent of them in the most contaminated fields (Section~\ref{subsec:comparison}) --- and these membership differences propagate into measurable offsets in the derived parameters. The fit-reliability framework of Section~\ref{sec:fit_reliability} then shows that a single, uniformly-applied pipeline reproduces the observed sequence of all 14 systems, and that where the ages are less precise the limitation lies in the information content of the CMDs themselves --- sparsely-populated turn-offs and differential reddening --- rather than in any failure of the fits, so that each parameter can be quoted with a quantified, cluster-specific confidence. The implication reaches beyond this sample: population-level studies that draw on automated cluster catalogues may be systematically biased in exactly this reddened, inner-disc regime, and the bias will be hardest to detect where the census looks most complete.

The transferable product of this work is therefore as much methodological as it is a parameter table. The diagnostics used here --- the width of the age posterior, the reduced chi-square $\chi^2_\nu$ of the fitted isochrone in the colour--magnitude plane, and the fraction $f_{1\sigma}$ of members within its $1\sigma$ contour --- are inexpensive, require no information beyond what the fit already produces, and together attach an explicit, per-cluster reliability label to each entry, a quantity that broader catalogues do not currently provide. Applied to other low-latitude clusters in a comparable observational situation, this offers a practical way of deciding which entries of a large catalogue can support population-level inference and which should be reserved for dedicated study. The natural next step is spectroscopic: multi-object observations with \textit{Gaia}-based membership pre-selection would anchor the chemical and radial-velocity profiles of these clusters and break the residual metallicity--reddening degeneracy that photometry alone cannot resolve, particularly for the distant and most heavily obscured systems.

In conclusion, this work modernises the legacy of Paris Pi\c{s}mi\c{s} by applying the precision of the \textit{Gaia} DR3 era to her historical discoveries. The clusters she found on the Tonantzintla plates were, by design, those hidden in the crowded low-latitude sky; that same property makes them, seven decades later, a stringent proving ground for the methods with which the modern census is built. In returning homogeneous parameters together with an explicit, quantitative statement of where those parameters can and cannot be trusted, this study aims to keep her discoveries usable as tracers of the Milky Way disc, and to offer a clear methodological path for the high-fidelity analysis of the many comparable systems that the large surveys have yet to reach.

\section*{Acknowledgements}
We thank the anonymous referee for a careful and constructive reading of the manuscript, which substantially improved this work. This study has been supported in part by the Scientific and Technological Research Council (T\"UB\.ITAK) 122F109. It was funded by the Scientific Research Projects Coordination Unit of Istanbul University (project number: 39743). This research has made use of the Astrophysics Data System, funded by NASA under Cooperative Agreement 80NSSC21M0056, as well as the SIMBAD and VizieR databases operated at CDS, Strasbourg, France, and presents results from the European Space Agency (ESA) space mission Gaia, whose data are processed by the Gaia Data Processing and Analysis Consortium (DPAC), funded by national institutions participating in the Gaia MultiLateral Agreement (MLA); the Gaia mission and archive websites are https://www.cosmos.esa.int/gaia and https://archives.esac.esa.int/gaia, respectively.

\section*{Data Availability}

All data underlying this work are either public or provided with the article. The photometry and astrometry are drawn from \textit{Gaia} DR3 \citep{Gaia_DR3} and are publicly available through the \textit{Gaia} Archive\footnote{\url{https://gea.esac.esa.int/archive/}}; the queries and quality cuts needed to reproduce our input catalogues are specified in full in Section~\ref{sec:query}. The theoretical isochrones were retrieved from the PARSEC CMD~3.9 web interface\footnote{\url{http://stev.oapd.inaf.it/cgi-bin/cmd}} with the grid parameters listed in Section~\ref{sec:method}. The fundamental parameters derived here, together with their posterior uncertainties, are tabulated in Table~\ref{tab:ns_results} and Table~\ref{tab:fundamental_results}, and the fit-reliability diagnostics in Table~\ref{tab:fit_reliability}. The membership catalogues for the 14 clusters --- containing positions, proper motions, parallaxes, \textit{Gaia} photometry and \textsc{UPMASK} membership probabilities for every star --- and the complete set of posterior and colour--magnitude diagnostic figures are provided as an additional figure and data set accompanying the online version of this article. Further material is available from the corresponding author upon reasonable request.

\clearpage

\appendix

\section{Fundamental Parameters of the Pismis OCs}
\label{app:fundamental}
Table~\ref{tab:fundamental_results} lists the astrometric parameters and Bayesian nested-sampling results for all 14 Pismis clusters analysed in this work.

The quantities in the table are of two distinct kinds, and their uncertainties are correspondingly of two kinds. The astrometric entries --- the mean proper-motion components and the mean parallax --- are averages over the member sample defined in Section~\ref{sec:membership}, and the quoted uncertainty is the uncertainty on that mean rather than the dispersion of the members about it. The parallax-based distance is obtained by direct inversion of the mean parallax, $d_\varpi = 1000/\varpi$ with $\varpi$ in mas, and its uncertainty follows by propagation, $\sigma_{d} = (d_\varpi^{2}/1000)\,\sigma_\varpi$; the reader should note that this propagation is asymmetric in reality and becomes increasingly so for the two most distant systems, Pismis~1 and Pismis~7, where $\sigma_\varpi/\varpi$ approaches $0.2$. The remaining entries --- reddening, initial metallicity, $[\mathrm{Fe/H}]$, $\log(\mathrm{Age})$, and the isochrone distance --- are posterior quantities from the nested-sampling run described in Section~\ref{sec:method}: the tabulated value is the posterior median and the asymmetric uncertainties are the 16th and 84th percentiles of the marginal posterior, so that they carry the full non-Gaussian shape of the distribution rather than a symmetric approximation to it. They are not independent of one another;  the correlations between age, distance, and reddening are shown explicitly in the posterior and CMD diagnostic figures provided as online supplementary material (see Data Availability) and discussed in Section~\ref{sec:fit_reliability}.

No zero-point correction has been applied to the \textit{Gaia} parallaxes, either to $d_\varpi$ or in the distance prior of the isochrone fit. The correction of \citet{Lindegren2021} is a function of magnitude, colour, and ecliptic latitude, and its global mean value, $\sim\!-0.02$~mas, is small compared with the mean parallaxes of most of our clusters but not negligible for the two most distant ones: for Pismis~1 and Pismis~7 ($\varpi \approx 0.18$~mas) it would shift the inverted distance by roughly $10\%$. This limitation applies to $d_\varpi$ as a reference quantity and, through the distance prior, to the isochrone-fit distances; it does not affect the age and reddening determinations at a comparable level, since these are driven by the morphology of the CMD rather than by its absolute vertical placement. A comparison of the parameters listed here with the values reported in the literature for each cluster is given, system by system, in Section~\ref{sec:individual_clusters}, and in aggregate against the major automated catalogues in Section~\ref{subsec:comparison}.

\clearpage
\renewcommand{\thetable}{A\arabic{table}}
\setcounter{table}{0}

\newsavebox\mybox
\newlength\mylength
\newcommand\boxup[2]{%
  \savebox\mybox{#1}%
  \setlength\mylength{\wd\mybox}%
  \parbox{\mylength}{#1 \\[1pt] #2}%
}

\begin{table*}[!ht]
\centering
\caption{Fundamental parameters of Pismis OCs derived from Bayesian nested-sampling analysis.}
\label{tab:fundamental_results}
\setlength{\tabcolsep}{7pt}
\renewcommand{\arraystretch}{1.1}
\footnotesize

\begin{tabular}{lccccc}
\hline
Cluster Name                               & Pismis 1                        & Pismis 3                        & Pismis 4                        & Pismis 5                        & Pismis 6                        \\[2pt]
\hline
$(\alpha, \delta)_{\rm J2000}$             & \boxup{08:18:18.8}{-37:06:06}   & \boxup{08:31:20.4}{-38:38:16}   & \boxup{08:35:09.6}{-44:25:26}   & \boxup{08:37:37.1}{-39:35:23}   & \boxup{08:39:07.9}{-46:13:36}   \\
$(l, b)_{\rm J2000}$ ($\deg$)              & 255.115,\,$-0.710$              & 257.852,\,$0.505$               & 262.929,\,$-2.363$              & 259.343,\,$0.912$               & 264.790,\,$-2.893$              \\
$\mu_{\alpha} \cos \delta$ (mas yr$^{-1}$) & $-2.571\pm$0.005                & $-4.775\pm$0.002                & $-8.186\pm$0.007                & $-5.509\pm$0.011                & $-5.883\pm$0.003                \\
$\mu_{\delta}$ (mas yr$^{-1}$)             & $3.952\pm$0.006                 & $6.705\pm$0.003                 & $5.343\pm$0.007                 & $4.297\pm$0.014                 & $5.125\pm$0.004                 \\
$\varpi$ (mas)                             & $0.180\pm$0.003                 & $0.437\pm$0.003                 & $1.459\pm$0.006                 & $1.058\pm$0.011                 & $0.550\pm$0.004                 \\
$d_{\varpi}$ (pc)                          & $5556\pm$93                     & $2288\pm$16                     & $685\pm$3                       & $945\pm$10                      & $1818\pm$13                     \\

$E(G_{\rm BP} - G_{\rm RP})$ (mag)         & $0.955^{+0.040}_{-0.045}$       & $1.575^{+0.142}_{-0.143}$       & $0.265^{+0.246}_{-0.143}$       & $0.600^{+0.253}_{-0.250}$       & $0.736^{+0.162}_{-0.119}$       \\
$Z_{\rm ini}$                              & $0.0092^{+0.0045}_{-0.0033}$    & $0.0111^{+0.0066}_{-0.0043}$    & $0.0084^{+0.0047}_{-0.0033}$    & $0.0098^{+0.0052}_{-0.0038}$    & $0.0100^{+0.0050}_{-0.0037}$    \\
${\rm [Fe/H]}$ (dex)                       & $-0.22^{+0.17}_{-0.19}$         & $-0.14^{+0.20}_{-0.21}$         & $-0.26^{+0.19}_{-0.21}$         & $-0.19^{+0.19}_{-0.22}$         & $-0.18^{+0.18}_{-0.20}$         \\ 
$\log(\mathrm{Age/yr})$                    & $7.75^{+0.20}_{-0.21}$          & $9.04^{+0.18}_{-0.21}$          & $8.35^{+0.55}_{-0.50}$          & $7.04^{+0.50}_{-0.48}$          & $7.88^{+0.41}_{-0.55}$          \\
Isochrone distance (pc)                    & $5445^{+14}_{-15}$              & $2293^{+15}_{-15}$              & $690^{+15}_{-15}$               & $926^{+14}_{-15}$               & $1815^{+15}_{-15}$              \\
\hline
\end{tabular}

\vspace{0.6cm}

\begin{tabular}{lccccc}
\hline
Cluster Name                               & Pismis 7                        & Pismis 8                        & Pismis 9                        & Pismis 12                       & Pismis 13                       \\[2pt]
\hline
$(\alpha, \delta)_{\rm J2000}$             & \boxup{08:41:08.2}{-38:41:58}   & \boxup{08:41:36.2}{-46:16:01}   & \boxup{08:42:56.2}{-44:55:55}   & \boxup{09:20:01.4}{-45:07:57}   & \boxup{09:22:07.0}{-51:06:06}   \\
$(l, b)_{\rm J2000}$ ($\deg$)              & 259.052,\,$1.995$               & 265.083,\,$-2.579$              & 264.173,\,$-1.574$              & 268.654,\,$3.208$               & 273.124,\,$-0.765$              \\
$\mu_{\alpha} \cos \delta$ (mas yr$^{-1}$) & $-3.287\pm$0.071                & $-5.667\pm$0.005                & $-4.383\pm$0.002                & $-6.717\pm$0.003                & $-5.087\pm$0.071                \\
$\mu_{\delta}$ (mas yr$^{-1}$)             & $2.830\pm$0.086                 & $4.973\pm$0.005                 & $3.021\pm$0.002                 & $4.891\pm$0.003                 & $4.668\pm$0.074                 \\
$\varpi$ (mas)                             & $0.184\pm$0.030                 & $0.523\pm$0.004                 & $0.505\pm$0.002                 & $0.447\pm$0.003                 & $0.349\pm$0.046                 \\
$d_{\varpi}$ (pc)                          & $5435\pm$886                    & $1912\pm$15                     & $1980\pm$8                      & $2237\pm$15                     & $2865\pm$378                    \\

$E(G_{\rm BP} - G_{\rm RP})$ (mag)         & $1.075^{+0.085}_{-0.211}$       & $0.981^{+0.167}_{-0.148}$       & $0.928^{+0.134}_{-0.136}$       & $0.798^{+0.146}_{-0.126}$       & $0.919^{+0.091}_{-0.114}$       \\
$Z_{\rm ini}$                              & $0.0092^{+0.0054}_{-0.0034}$    & $0.0108^{+0.0056}_{-0.0043}$    & $0.0111^{+0.0064}_{-0.0044}$    & $0.0104^{+0.0049}_{-0.0036}$    & $0.0176^{+0.0052}_{-0.0061}$    \\
${\rm [Fe/H]}$ (dex)                       & $-0.22^{+0.20}_{-0.20}$         & $-0.15^{+0.18}_{-0.22}$         & $-0.14^{+0.20}_{-0.22}$         & $-0.17^{+0.17}_{-0.19}$         & $+0.06^{+0.11}_{-0.19}$         \\ 
$\log(\mathrm{Age/yr})$                    & $8.64^{+0.32}_{-0.11}$          & $8.28^{+0.43}_{-0.53}$          & $8.30^{+0.32}_{-0.46}$          & $9.18^{+0.30}_{-0.17}$          & $8.35^{+0.26}_{-0.30}$          \\
Isochrone distance (pc)                    & $5367^{+15}_{-15}$              & $1865^{+16}_{-15}$              & $1995^{+16}_{-14}$              & $2233^{+15}_{-15}$              & $2900^{+15}_{-15}$              \\
\hline
\end{tabular}

\vspace{0.6cm}

\begin{tabular}{lcccc}
\hline
Cluster Name                               & Pismis 14                       & Pismis 15                       & Pismis 18                       & Pismis 19                       \\[2pt]
\hline
$(\alpha, \delta)_{\rm J2000}$             & \boxup{09:29:52.8}{-52:46:48}   & \boxup{09:34:44.7}{-48:02:36}   & \boxup{13:36:54.8}{-62:05:29}   & \boxup{14:30:40.0}{-60:53:21}   \\
$(l, b)_{\rm J2000}$ ($\deg$)              & 275.151,\,$-1.139$              & 272.499,\,$2.861$               & 308.226,\,$0.312$               & 314.710,\,$-0.307$              \\
$\mu_{\alpha} \cos \delta$ (mas yr$^{-1}$) & $-6.755\pm$0.043                & $-5.248\pm$0.098                & $-5.720\pm$0.074                & $-5.551\pm$0.003                \\
$\mu_{\delta}$ (mas yr$^{-1}$)             & $6.152\pm$0.038                 & $3.366\pm$0.077                 & $-2.241\pm$0.056                & $-3.336\pm$0.003                \\
$\varpi$ (mas)                             & $0.786\pm$0.055                 & $0.466\pm$0.075                 & $0.478\pm$0.041                 & $0.408\pm$0.002                 \\
$d_{\varpi}$ (pc)                          & $1272\pm$89                     & $2146\pm$345                    & $2092\pm$179                    & $2451\pm$12                     \\

$E(G_{\rm BP} - G_{\rm RP})$ (mag)         & $0.442^{+0.119}_{-0.107}$       & $0.969^{+0.155}_{-0.165}$       & $1.160^{+0.135}_{-0.187}$       & $0.799^{+0.189}_{-0.340}$       \\
$Z_{\rm ini}$                              & $0.0097^{+0.0048}_{-0.0039}$    & $0.0124^{+0.0062}_{-0.0044}$    & $0.0143^{+0.0060}_{-0.0060}$    & $0.0103^{+0.0060}_{-0.0041}$    \\
${\rm [Fe/H]}$ (dex)                       & $-0.20^{+0.17}_{-0.22}$         & $-0.09^{+0.18}_{-0.19}$         & $-0.03^{+0.15}_{-0.24}$         & $-0.17^{+0.20}_{-0.22}$         \\ 
$\log(\mathrm{Age/yr})$                    & $8.78^{+0.36}_{-0.39}$          & $9.09^{+0.35}_{-0.21}$          & $8.54^{+0.32}_{-0.43}$          & $7.69^{+0.52}_{-0.55}$          \\
Isochrone distance (pc)                    & $1291^{+15}_{-14}$              & $2421^{+14}_{-16}$              & $2807^{+15}_{-14}$              & $2400^{+14}_{-15}$              \\
\hline
\end{tabular}
\end{table*}

\bibliographystyle{aasjournalv7}
\bibliography{references_merged}

\begin{thebibliography}{}
\expandafter\ifx\csname natexlab\endcsname\relax\def\natexlab#1{#1}\fi
\providecommand{\url}[1]{\href{#1}{#1}}
\providecommand{\dodoi}[1]{doi:~\href{http://doi.org/#1}{\nolinkurl{#1}}}
\providecommand{\doeprint}[1]{\href{http://ascl.net/#1}{\nolinkurl{http://ascl.net/#1}}}
\providecommand{\doarXiv}[1]{\href{https://arxiv.org/abs/#1}{\nolinkurl{https://arxiv.org/abs/#1}}}

\bibitem[{N.~M. {Ahmed} \& A.~L. {Tadross}(2025){Ahmed} \&
  {Tadross}}]{Nasser_2025}
{Ahmed}, N.~M., \& {Tadross}, A.~L. 2025, \bibinfo{title}{{Precision analysis
  of NGC 2158 with Gaia DR3},} Scientific Reports, 15, 20189,
  \dodoi{10.1038/s41598-025-06119-1}

\bibitem[{J.~A. {Ahumada}(2005){Ahumada}}]{Ahumada_2005}
{Ahumada}, J.~A. 2005, \bibinfo{title}{{CCD photometry of the open clusters NGC
  2627, NGC 5617, Pismis 7, and Ruprecht 75},} Astronomische Nachrichten, 326,
  3, \dodoi{10.1002/asna.200310257}

\bibitem[{F. {Akbaba} {et~al.}(2024){Akbaba}, {Ak}, {Bilir}, {Plevne},
  {{\~A}-nal Ta{\c{s}}}, \& {Seabroke}}]{Akbaba2024}
{Akbaba}, F., {Ak}, T., {Bilir}, S., {et~al.} 2024, \bibinfo{title}{{Radial
  Metallicity Gradients for the Chemically Selected Galactic Thin Disc
  Main-Sequence Stars},} Astronomische Nachrichten, 345, e20240052,
  \dodoi{10.1002/asna.20240052}

\bibitem[{F. {Akbaba} {et~al.}(2026){Akbaba}, {Plevne}, {{\c{S}}ahin}, \&
  {{\c{S}}ent{\"u}rk}}]{Akbaba2026}
{Akbaba}, F., {Plevne}, O., {{\c{S}}ahin}, T., \& {{\c{S}}ent{\"u}rk}, S.~A.
  2026, \bibinfo{title}{{Chemical taxonomy of {\ensuremath{\omega}} Centauri:
  ten populations reveal a multiphase enrichment history},} \mnras, 550,
  stag1308, \dodoi{10.1093/mnras/stag1308}

\bibitem[{R. {Andrae} {et~al.}(2023){Andrae}, {Rix}, \& {Chandra}}]{Andrae2023}
{Andrae}, R., {Rix}, H.-W., \& {Chandra}, V. 2023, \bibinfo{title}{{Robust
  Data-driven Metallicities for 175 Million Stars from Gaia XP Spectra},}
  \apjs, 267, 8, \dodoi{10.3847/1538-4365/acd53e}

\bibitem[{C.~A.~L. {Bailer-Jones} {et~al.}(2021){Bailer-Jones}, {Rybizki},
  {Fouesneau}, {Demleitner}, \& {Andrae}}]{Bailer-Jones21}
{Bailer-Jones}, C.~A.~L., {Rybizki}, J., {Fouesneau}, M., {Demleitner}, M., \&
  {Andrae}, R. 2021, \bibinfo{title}{{Estimating Distances from Parallaxes. V.
  Geometric and Photogeometric Distances to 1.47 Billion Stars in Gaia Early
  Data Release 3},} \aj, 161, 147, \dodoi{10.3847/1538-3881/abd806}

\bibitem[{L.~D. {Baravalle} {et~al.}(2021){Baravalle}, {Alonso}, {Minniti},
  {Nilo Castell{\'o}n}, {Soto}, {Valotto}, {Villal{\'o}n}, {Gra{\~n}a},
  {Am{\^o}res}, \& {Milla Castro}}]{Baravalle2021}
{Baravalle}, L.~D., {Alonso}, M.~V., {Minniti}, D., {et~al.} 2021,
  \bibinfo{title}{{The VVV near-IR galaxy catalogue beyond the Galactic disc},}
  \mnras, 502, 601, \dodoi{10.1093/mnras/staa4020}

\bibitem[{K. {Belwal} {et~al.}(2025){Belwal}, {Bisht}, {Jiang}, {Yadav}, {Raj},
  {Rangwal}, {Dattatrey}, {Bisht}, \& {Durgapal}}]{Belwal_2025}
{Belwal}, K., {Bisht}, D., {Jiang}, I.-G., {et~al.} 2025,
  \bibinfo{title}{{Unveiling dynamics and variability in open clusters:
  insights from a comprehensive analysis of six galactic clusters},} \mnras,
  544, 988, \dodoi{10.1093/mnras/staf1739}

\bibitem[{E. {Bica} \& C. {Bonatto}(2011){Bica} \& {Bonatto}}]{Bica_2011}
{Bica}, E., \& {Bonatto}, C. 2011, \bibinfo{title}{{Star clusters or asterisms?
  2MASS CMD and structural analyses of 15 challenging targets},} \aap, 530,
  A32, \dodoi{10.1051/0004-6361/201116452}

\bibitem[{D. {Bisht} {et~al.}(2022){Bisht}, {Zhu}, {Elsanhoury}, {Yadav},
  {Rangwal}, {Sariya}, {Durgapal}, \& {Jiang}}]{Bisht2022}
{Bisht}, D., {Zhu}, Q., {Elsanhoury}, W.~H., {et~al.} 2022, \bibinfo{title}{{A
  Comprehensive Study of Five Intermediate-age Pismis (2, 3, 7, 12, 15)
  Clusters Using Photometric and Astrometric Data from Gaia EDR3},} \aj, 164,
  171, \dodoi{10.3847/1538-3881/ac8cf4}

\bibitem[{C. {Bonatto} \& E. {Bica}(2009){Bonatto} \& {Bica}}]{Bonatto_2009}
{Bonatto}, C., \& {Bica}, E. 2009, \bibinfo{title}{{The nature of the young and
  low-mass open clusters Pismis 5, vdB 80, NGC 1931 and BDSB 96},} \mnras, 397,
  1915, \dodoi{10.1111/j.1365-2966.2009.14877.x}

\bibitem[{D. {Bossini} {et~al.}(2019){Bossini}, {Vallenari}, {Bragaglia},
  {Cantat-Gaudin}, {Sordo}, {Balaguer-N{\'u}{\~n}ez}, {Jordi}, {Moitinho},
  {Soubiran}, {Casamiquela}, {Carrera}, \& {Heiter}}]{Bossini_2019}
{Bossini}, D., {Vallenari}, A., {Bragaglia}, A., {et~al.} 2019,
  \bibinfo{title}{{Age determination for 269 Gaia DR2 open clusters},} \aap,
  623, A108, \dodoi{10.1051/0004-6361/201834693}

\bibitem[{J. {Bovy} {et~al.}(2016){Bovy}, {Rix}, {Green}, {Schlafly}, \&
  {Finkbeiner}}]{mwdust}
{Bovy}, J., {Rix}, H.-W., {Green}, G.~M., {Schlafly}, E.~F., \& {Finkbeiner},
  D.~P. 2016, \bibinfo{title}{{On Galactic Density Modeling in the Presence of
  Dust Extinction},} \apj, 818, 130, \dodoi{10.3847/0004-637X/818/2/130}

\bibitem[{A. Bressan {et~al.}(2012)Bressan, Marigo, Girardi, Salasnich,
  Dal~Cero, Rubele, \& Nanni}]{Bressan_2012}
Bressan, A., Marigo, P., Girardi, L., {et~al.} 2012, \bibinfo{title}{{PARSEC}:
  stellar tracks and isochrones with the {PA}dova and {TR}ieste {S}tellar
  {E}volution {C}ode,} \mnras, 427, 127,
  \dodoi{10.1111/j.1365-2966.2012.21948.x}

\bibitem[{S. {Buder} {et~al.}(2025){Buder}, {Kos}, {Wang}, {McKenzie},
  {Howell}, {Martell}, {Hayden}, {Zucker}, {Nordlander}, {Montet}, {Traven},
  {Bland-Hawthorn}, {de Silva}, {Freeman}, {Lewis}, {Lind}, {Sharma},
  {Simpson}, {Stello}, {Zwitter}, {Amarsi}, {Armstrong}, {Banks}, {Beavis},
  {Beeson}, {Chen}, {Ciuc{\u{a}}}, {da Costa}, {de Grijs}, {Martin}, {Nataf},
  {Ness}, {Rains}, {Scarr}, {Vogrin{\v{c}}i{\v{c}}}, {Wang}, {Wittenmyer},
  {Xie}, \& {The Galah Collaboration}}]{Buder2025}
{Buder}, S., {Kos}, J., {Wang}, X.~E., {et~al.} 2025, \bibinfo{title}{{The
  GALAH survey: Data release 4},} \pasa, 42, e051, \dodoi{10.1017/pasa.2025.26}

\bibitem[{{\L}. {Bukowiecki} {et~al.}(2011){Bukowiecki}, {Maciejewski},
  {Konorski}, \& {Strobel}}]{Bukowiecki_2011}
{Bukowiecki}, {\L}., {Maciejewski}, G., {Konorski}, P., \& {Strobel}, A. 2011,
  \bibinfo{title}{{Open Clusters in 2MASS Photometry. I. Structural and Basic
  Astrophysical Parameters},} \actaa, 61, 231, \dodoi{10.48550/arXiv.1107.5119}

\bibitem[{T. Cantat-Gaudin {et~al.}(2018)Cantat-Gaudin, Jordi, Vallenari,
  Bragaglia, Balaguer-N{\'u}{\~n}ez, Soubiran, Bossini, Moitinho,
  Castro-Ginard, Krone-Martins, Casamiquela, Sordo, \&
  Carrera}]{Cantat-Gaudin_2018}
Cantat-Gaudin, T., Jordi, C., Vallenari, A., {et~al.} 2018, \bibinfo{title}{{A
  {G}aia {DR2} view of the open cluster population in the {M}ilky {W}ay},}
  \aap, 618, A93, \dodoi{10.1051/0004-6361/201833476}

\bibitem[{T. Cantat-Gaudin {et~al.}(2020)Cantat-Gaudin, Anders, Castro-Ginard,
  Jordi, Romero-G{\'o}mez, Soubiran, Casamiquela, Tarricq, Moitinho, Vallenari,
  Bragaglia, Krone-Martins, \& Kounkel}]{Cantat-Gaudin_2020}
Cantat-Gaudin, T., Anders, F., Castro-Ginard, A., {et~al.} 2020,
  \bibinfo{title}{Painting a portrait of the Galactic disc with its stellar
  clusters,} \aap, 640, A1, \dodoi{10.1051/0004-6361/202038192}

\bibitem[{G. {Carraro} {et~al.}(2005){Carraro}, {Geisler}, {Baume},
  {V{\'a}zquez}, \& {Moitinho}}]{Carraro_2005}
{Carraro}, G., {Geisler}, D., {Baume}, G., {V{\'a}zquez}, R., \& {Moitinho}, A.
  2005, \bibinfo{title}{{The intermediate-age open clusters Ruprecht 4,
  Ruprecht 7 and Pismis 15},} \mnras, 360, 655,
  \dodoi{10.1111/j.1365-2966.2005.09054.x}

\bibitem[{G. {Carraro} \& U. {Munari}(2004){Carraro} \&
  {Munari}}]{Carraro_2004}
{Carraro}, G., \& {Munari}, U. 2004, \bibinfo{title}{{A multicolour CCD
  photometric study of the open clusters NGC 2866, Pismis 19, Westerlund 2,
  ESO96-SC04, NGC 5617 and NGC 6204},} \mnras, 347, 625,
  \dodoi{10.1111/j.1365-2966.2004.07244.x}

\bibitem[{G. {Carraro} \& S. {Ortolani}(1994){Carraro} \&
  {Ortolani}}]{Carraro_1994}
{Carraro}, G., \& {Ortolani}, S. 1994, \bibinfo{title}{{PISMIS 3: an unstudied
  old open cluster.},} \aap, 291, 106

\bibitem[{G. {Carraro} {et~al.}(2017){Carraro}, {Sales Silva}, {Moni Bidin}, \&
  {Vazquez}}]{Carraro_2017}
{Carraro}, G., {Sales Silva}, J.~V., {Moni Bidin}, C., \& {Vazquez}, R.~A.
  2017, \bibinfo{title}{{Galactic Structure in the Outer Disk: The Field in the
  Line of Sight to the Intermediate-age Open Cluster Tombaugh 1},} \aj, 153,
  99, \dodoi{10.3847/1538-3881/153/3/99}

\bibitem[{G. {Carraro} \& A.~F. {Seleznev}(2012){Carraro} \&
  {Seleznev}}]{Carraro2012}
{Carraro}, G., \& {Seleznev}, A.~F. 2012, \bibinfo{title}{{UBVI CCD photometry
  and star counts in nine inner disc Galactic star clusters},} \mnras, 419,
  3608, \dodoi{10.1111/j.1365-2966.2011.20010.x}

\bibitem[{A. {Castro-Ginard} {et~al.}(2020){Castro-Ginard}, {Jordi}, {Luri},
  {{\'A}lvarez Cid-Fuentes}, {Casamiquela}, {Anders}, {Cantat-Gaudin},
  {Mongui{\'o}}, {Balaguer-N{\'u}{\~n}ez}, {Sol{\`a}}, \&
  {Badia}}]{Castro-Ginard2020}
{Castro-Ginard}, A., {Jordi}, C., {Luri}, X., {et~al.} 2020,
  \bibinfo{title}{Hunting for open clusters in {G}aia {DR}2: 582 new open
  clusters in the {G}alactic disc,} \aap, 635, A45,
  \dodoi{10.1051/0004-6361/201937386}

\bibitem[{A. {Castro-Ginard} {et~al.}(2022){Castro-Ginard}, {Jordi}, {Luri},
  {Cantat-Gaudin}, {Carrasco}, {Casamiquela}, {Anders},
  {Balaguer-N{\'u}{\~n}ez}, \& {Badia}}]{Castro-Ginard2022}
{Castro-Ginard}, A., {Jordi}, C., {Luri}, X., {et~al.} 2022,
  \bibinfo{title}{{Hunting for open clusters in Gaia EDR3: 628 new open
  clusters found with OCfinder},} \aap, 661, A118,
  \dodoi{10.1051/0004-6361/202142568}

\bibitem[{L. {Cavallo} {et~al.}(2024){Cavallo}, {Spina}, {Carraro}, {Magrini},
  {Poggio}, {Cantat-Gaudin}, {Pasquato}, {Lucatello}, {Ortolani}, \&
  {Schiappacasse-Ulloa}}]{Cavallo_2024}
{Cavallo}, L., {Spina}, L., {Carraro}, G., {et~al.} 2024,
  \bibinfo{title}{{Parameter Estimation for Open Clusters using an Artificial
  Neural Network with a QuadTree-based Feature Extractor},} \aj, 167, 12,
  \dodoi{10.3847/1538-3881/ad07e5}

\bibitem[{H. {{\c{C}}akmak} {et~al.}(2021){{\c{C}}akmak},
  {G{\"u}ne{\textcommabelow s}}, {Karata{\textcommabelow s}}, \&
  {Bonatto}}]{Cakmak_2021}
{{\c{C}}akmak}, H., {G{\"u}ne{\textcommabelow s}}, O., {Karata{\textcommabelow
  s}}, Y., \& {Bonatto}, C. 2021, \bibinfo{title}{{Astrophysical parameters and
  dynamical evolution of open clusters: NGC 2587, Collinder 268 and Melotte 72,
  and Pismis 7},} Astronomische Nachrichten, 342, 975,
  \dodoi{10.1002/asna.202113983}

\bibitem[{H. {{\c{C}}akmak} {et~al.}(2024){{\c{C}}akmak}, {Yontan}, {Bilir},
  {Banks}, {Michel}, {Soydugan}, {Ko{\c{c}}}, \& {Er{\c{c}}ay}}]{Cakmak_2024}
{{\c{C}}akmak}, H., {Yontan}, T., {Bilir}, S., {et~al.} 2024,
  \bibinfo{title}{{Photometric and Kinematic Studies of Open Clusters Ruprecht
  1 and Ruprecht 171},} Astronomische Nachrichten, 345, e20240054,
  \dodoi{10.1002/asna.20240054}

\bibitem[{J.~J. {Claria}(1979){Claria}}]{Claria_1979}
{Claria}, J.~J. 1979, \bibinfo{title}{{PISMIS 13: a small, very compact open
  cluster in VELA},} The Observatory, 99, 202

\bibitem[{W.~S. {Dias} {et~al.}(2002){Dias}, {Alessi}, {Moitinho}, \&
  {L{\'e}pine}}]{Dias_2002}
{Dias}, W.~S., {Alessi}, B.~S., {Moitinho}, A., \& {L{\'e}pine}, J.~R.~D. 2002,
  \bibinfo{title}{{New catalogue of optically visible open clusters and
  candidates},} \aap, 389, 871, \dodoi{10.1051/0004-6361:20020668}

\bibitem[{W.~S. Dias {et~al.}(2021)Dias, Monteir, Moitinho, L{\'e}pine,
  Carraro, Paunzen, Alessi, \& Villela}]{Dias_2021}
Dias, W.~S., Monteir, H., Moitinho, A., {et~al.} 2021, \bibinfo{title}{Updated
  parameters of 1743 open clusters based on {G}aia {DR2},} \mnras, 504, 356,
  \dodoi{10.1093/mnras/stab770}

\bibitem[{J. Donor {et~al.}(2020)Donor, Frinchaboy, Cunha, OâConnell,
  Prieto, Almeida, Anders, Beaton, Bizyaev, Brownstein, Carrera, Chiappini,
  Cohen, GarcÃ­a-HernÃ¡ndez, Geisler, Hasselquist, JÃ¶nsson, Lane,
  Majewski, Minniti, Bidin, Pan, Roman-Lopes, Sobeck, \& Zasowski}]{Donor_2020}
Donor, J., Frinchaboy, P.~M., Cunha, K., {et~al.} 2020, \bibinfo{title}{The
  Open Cluster Chemical Abundances and Mapping Survey. {IV.} Abundances for 128
  Open Clusters Using {SDSS/APOGEE} {DR}16,} The Astronomical Journal, 159,
  199, \dodoi{10.3847/1538-3881/ab77bc}

\bibitem[{M.~P. {Fitzgerald} {et~al.}(1979){Fitzgerald}, {Boudreault}, {Fich},
  {Luiken}, \& {Witt}}]{fitzgerald1979}
{Fitzgerald}, M.~P., {Boudreault}, R., {Fich}, M., {Luiken}, M., \& {Witt},
  A.~N. 1979, \bibinfo{title}{{The open clusters Pismis 6 and 8, and Wat 6.},}
  \aaps, 37, 351

\bibitem[{M.~P. {FitzGerald} {et~al.}(1979){FitzGerald}, {Luiken}, {Maitzen},
  \& {Moffat}}]{FitzGerald_1979}
{FitzGerald}, M.~P., {Luiken}, M., {Maitzen}, H.~M., \& {Moffat}, A. F.~J.
  1979, \bibinfo{title}{{UBV photometry of the open clusters Pismis 8, Pismis
  13, and NGC 2627},} \aaps, 37, 345

\bibitem[{D. {Forbes} \& S. {Short}(1994){Forbes} \& {Short}}]{Forbes_1994}
{Forbes}, D., \& {Short}, S. 1994, \bibinfo{title}{{PISMIS 6 and the FO IA
  Supergiant HD 74180},} \aj, 108, 594, \dodoi{10.1086/117092}

\bibitem[{E.~D. {Friel}(1995){Friel}}]{Friel1995}
{Friel}, E.~D. 1995, \bibinfo{title}{{The Old Open Clusters of the Milky Way},}
  \araa, 33, 381, \dodoi{10.1146/annurev.aa.33.090195.002121}

\bibitem[{ {Gaia Collaboration} {et~al.}(2016){Gaia Collaboration}, {Prusti},
  {de Bruijne}, {et~al.}}]{Gaia16}
{Gaia Collaboration}, {Prusti}, T., {de Bruijne}, J.~H.~J., {et~al.} 2016,
  \bibinfo{title}{{The Gaia mission},} Astronomy \& Astrophysics, 595, A1,
  \dodoi{10.1051/0004-6361/201629272}

\bibitem[{ {Gaia Collaboration} {et~al.}(2023){Gaia Collaboration},
  {Vallenari}, {Brown}, {et~al.}}]{Gaia_DR3}
{Gaia Collaboration}, {Vallenari}, A., {Brown}, A.~G.~A., {et~al.} 2023,
  \bibinfo{title}{{Gaia Data Release 3. Summary of the content and survey
  properties},} \aap, 674, A1, \dodoi{10.1051/0004-6361/202243940}

\bibitem[{E.~E. {Giorgi} {et~al.}(2005){Giorgi}, {Baume}, {Solivella}, \&
  {V{\'a}zquez}}]{Giorgi_2005}
{Giorgi}, E.~E., {Baume}, G., {Solivella}, G., \& {V{\'a}zquez}, R.~A. 2005,
  \bibinfo{title}{{The Vela-Puppis open clusters Pismis 8 and Pismis 13},}
  \aap, 432, 491, \dodoi{10.1051/0004-6361:20041075}

\bibitem[{E.~E. {Giorgi} {et~al.}(2023){Giorgi}, {Pera}, {Perren}, {Vazquez},
  \& {Cruzado}}]{Giorgi_2023}
{Giorgi}, E.~E., {Pera}, M.~S., {Perren}, G.~I., {Vazquez}, R.~A., \&
  {Cruzado}, A. 2023, \bibinfo{title}{{NGC 2659: a probable binary cluster},}
  Boletin de la Asociacion Argentina de Astronomia La Plata Argentina, 64, 90

\bibitem[{J. {Gregorio-Hetem} \& A. {Hetem}(2024){Gregorio-Hetem} \&
  {Hetem}}]{Hetem_2024}
{Gregorio-Hetem}, J., \& {Hetem}, A. 2024, \bibinfo{title}{{Structural
  properties of subgroups of stars associated with open clusters},} \mnras,
  533, 1782, \dodoi{10.1093/mnras/stae1869}

\bibitem[{D. {Hatzidimitriou} {et~al.}(2019){Hatzidimitriou}, {Held},
  {Tognelli}, {et~al.}}]{Hatzidimitriou_2019}
{Hatzidimitriou}, D., {Held}, E.~V., {Tognelli}, E., {et~al.} 2019,
  \bibinfo{title}{{The Gaia-ESO Survey: The inner disc, intermediate-age open
  cluster Pismis 18},} \aap, 626, A90, \dodoi{10.1051/0004-6361/201834636}

\bibitem[{S. {Herschel}(1847){Herschel}}]{Herschel1847}
{Herschel}, John Frederick~William, S. 1847, {Results of astronomical
  observations made during the years 1834, 5, 6, 7, 8, at the Cape of Good
  Hope; being the completion of a telescopic survey of the whole surface of the
  visible heavens, commenced in 1825} ({Smith, Elder \& Co})

\bibitem[{E. {Higson} {et~al.}(2019){Higson}, {Handley}, {Hobson}, \&
  {Lasenby}}]{Higson2019}
{Higson}, E., {Handley}, W., {Hobson}, M., \& {Lasenby}, A. 2019,
  \bibinfo{title}{{Dynamic nested sampling: an improved algorithm for parameter
  estimation and evidence calculation},} Statistics and Computing, 29, 891,
  \dodoi{10.1007/s11222-018-9844-0}

\bibitem[{D. {Horta} {et~al.}(2026){Horta}, {Price-Whelan}, {Koposov}, {Hunt},
  {Hogg}, {Filion}, \& {Daniel}}]{Horta2026}
{Horta}, D., {Price-Whelan}, A.~M., {Koposov}, S.~E., {et~al.} 2026,
  \bibinfo{title}{{The Milky Way's Circular Velocity Curve Measured Using
  Element Abundance Gradients},} \apj, 1000, 8,
  \dodoi{10.3847/1538-4357/ae43de}

\bibitem[{E.~L. {Hunt} \& S. {Reffert}(2023){Hunt} \& {Reffert}}]{Hunt_2023}
{Hunt}, E.~L., \& {Reffert}, S. 2023, \bibinfo{title}{{Improving the open
  cluster census. II. An all-sky cluster catalogue with Gaia DR3},} \aap, 673,
  A114, \dodoi{10.1051/0004-6361/202346285}

\bibitem[{E.~L. {Hunt} \& S. {Reffert}(2024){Hunt} \& {Reffert}}]{Hunt_2024}
{Hunt}, E.~L., \& {Reffert}, S. 2024, \bibinfo{title}{{Improving the open
  cluster census. III. Using cluster masses, radii, and dynamics to create a
  cleaned open cluster catalogue},} \aap, 686, A42,
  \dodoi{10.1051/0004-6361/202348662}

\bibitem[{Y.~C. {Joshi} {et~al.}(2016){Joshi}, {Dambis}, {Pandey}, \&
  {Joshi}}]{Joshi_2016}
{Joshi}, Y.~C., {Dambis}, A.~K., {Pandey}, A.~K., \& {Joshi}, S. 2016,
  \bibinfo{title}{{Study of open clusters within 1.8 kpc and understanding the
  Galactic structure},} \aap, 593, A116, \dodoi{10.1051/0004-6361/201628944}

\bibitem[{H. {Karag{\"o}z} {et~al.}(2025){Karag{\"o}z}, {Yontan}, {Bilir},
  {Plevne}, {Ak}, {Ak}, {Canbay}, \& {Banks}}]{Karagoz_2025}
{Karag{\"o}z}, H., {Yontan}, T., {Bilir}, S., {et~al.} 2025, \bibinfo{title}{{A
  Multidata Approach to Open Clusters: Roslund 3 and Ruprecht 174 in CCD UBV
  and Gaia DR3 Context},} \aj, 170, 149, \dodoi{10.3847/1538-3881/adef16}

\bibitem[{N.~V. Kharchenko {et~al.}(2013)Kharchenko, Piskunov, Schilbach,
  R{\"o}ser, \& Scholz}]{Kharchenko_2013}
Kharchenko, N.~V., Piskunov, A.~E., Schilbach, E., R{\"o}ser, S., \& Scholz,
  R.~D. 2013, \bibinfo{title}{Global survey of star clusters in the {M}ilky
  {W}ay. {II.} The catalogue of basic parameters,} \aap, 558, A53,
  \dodoi{10.1051/0004-6361/201322302}

\bibitem[{S.~C. {Kim} {et~al.}(2017){Kim}, {Kyeong}, {Park}, {Han}, {Lee},
  {Moon}, {Lee}, \& {Kim}}]{Kim_2017}
{Kim}, S.~C., {Kyeong}, J., {Park}, H.~S., {et~al.} 2017, \bibinfo{title}{{BVI
  Photometric Study of the Old Open Cluster Ruprecht 6},} Journal of Korean
  Astronomical Society, 50, 79, \dodoi{10.5303/JKAS.2017.50.3.79}

\bibitem[{S. {Koc} \& T. {Yontan}(2023){Koc} \& {Yontan}}]{Koc_2023}
{Koc}, S., \& {Yontan}, T. 2023, \bibinfo{title}{{Astrophysical Parameters of
  the Open Cluster Berkeley 6},} arXiv e-prints, arXiv:2306.15367,
  \dodoi{10.48550/arXiv.2306.15367}

\bibitem[{J.~A. {Kollmeier} {et~al.}(2026){Kollmeier}, {Rix}, {Aerts},
  {et~al.}}]{Kollmeier2026}
{Kollmeier}, J.~A., {Rix}, H.-W., {Aerts}, C., {et~al.} 2026,
  \bibinfo{title}{{Sloan Digital Sky Survey. V. Pioneering Panoptic
  Spectroscopy},} \aj, 171, 52, \dodoi{10.3847/1538-3881/ae0576}

\bibitem[{V. {Kopchev} {et~al.}(2008){Kopchev}, {Nedialkov}, \&
  {Petrov}}]{Kopchev_2008}
{Kopchev}, V., {Nedialkov}, P., \& {Petrov}, G. 2008, \bibinfo{title}{{Age
  determination of possible binary open clusters NGC 2383/NGC 2384 and Pismis
  6/Pismis 8},} arXiv e-prints, arXiv:0808.4055,
  \dodoi{10.48550/arXiv.0808.4055}

\bibitem[{A. Krone-Martins \& A. Moitinho(2014)Krone-Martins \&
  Moitinho}]{Krone-Martins_2014}
Krone-Martins, A., \& Moitinho, A. 2014, \bibinfo{title}{{UPMASK}: unsupervised
  photometric membership assignment in stellar clusters,} \aap, 561, A57,
  \dodoi{10.1051/0004-6361/201321143}

\bibitem[{M.~R. Krumholz {et~al.}(2019)Krumholz, McKee, \&
  Bland-Hawthorn}]{Krumholz_2019}
Krumholz, M.~R., McKee, C.~F., \& Bland-Hawthorn, J. 2019, \bibinfo{title}{Star
  Clusters Across Cosmic Time,} Annual Review of Astronomy and Astrophysics,
  57, 227, \dodoi{10.1146/annurev-astro-091918-104430}

\bibitem[{C.~J. Lada \& E.~A. Lada(2003)Lada \& Lada}]{Lada_2003}
Lada, C.~J., \& Lada, E.~A. 2003, \bibinfo{title}{Embedded Clusters in
  Molecular Clouds,} \araa, 41, 57,
  \dodoi{10.1146/annurev.astro.41.011802.094844}

\bibitem[{J. {Li} {et~al.}(2024){Li}, {Wong}, {Hogg}, {Rix}, \&
  {Chandra}}]{Li2024}
{Li}, J., {Wong}, K. W.~K., {Hogg}, D.~W., {Rix}, H.-W., \& {Chandra}, V. 2024,
  \bibinfo{title}{{AspGap: Augmented Stellar Parameters and Abundances for 37
  Million Red Giant Branch Stars from Gaia XP Low-resolution Spectra},} \apjs,
  272, 2, \dodoi{10.3847/1538-4365/ad2b4d}

\bibitem[{L. {Lindegren} {et~al.}(2018){Lindegren}, {Hern{\'a}ndez}, {Bombrun},
  {et~al.}}]{Lindegren_2018}
{Lindegren}, L., {Hern{\'a}ndez}, J., {Bombrun}, A., {et~al.} 2018,
  \bibinfo{title}{{Gaia Data Release 2. The astrometric solution},} \aap, 616,
  A2, \dodoi{10.1051/0004-6361/201832727}

\bibitem[{L. {Lindegren} {et~al.}(2021){Lindegren}, {Klioner}, {Hern{\'a}ndez},
  {et~al.}}]{Lindegren2021}
{Lindegren}, L., {Klioner}, S.~A., {Hern{\'a}ndez}, J., {et~al.} 2021,
  \bibinfo{title}{{Gaia Early Data Release 3. The astrometric solution},} \aap,
  649, A2, \dodoi{10.1051/0004-6361/202039709}

\bibitem[{D. {Majaess} {et~al.}(2025){Majaess}, {Bonatto}, {Turner}, {Saito},
  {Minniti}, {Moni Bidin}, {Gonz{\'a}lez-D{\'\i}az}, {Alonso-Garcia}, {Bono},
  {Braga}, {Navarro}, {Carraro}, \& {Gomez}}]{Majaess_2025}
{Majaess}, D., {Bonatto}, C.~J., {Turner}, D.~G., {et~al.} 2025,
  \bibinfo{title}{{The Gaia Parallax Discrepancy for the Cluster Pismis 19 and
  Separating {\ensuremath{\delta}} Scutis from Cepheids},} \apj, 982, 165,
  \dodoi{10.3847/1538-4357/adb9e4}

\bibitem[{H. {Monteiro} {et~al.}(2021){Monteiro}, {Barros}, {Dias}, \&
  {L{\'e}pine}}]{Monteiro_2021}
{Monteiro}, H., {Barros}, D.~A., {Dias}, W.~S., \& {L{\'e}pine}, J. R.~D. 2021,
  \bibinfo{title}{{The distribution of open clusters in the Galaxy},} Frontiers
  in Astronomy and Space Sciences, 8, 62, \dodoi{10.3389/fspas.2021.656474}

\bibitem[{M. {Netopil} {et~al.}(2022){Netopil}, {Oralhan}, {{\c{C}}akmak},
  {Michel}, \& {Karata{\c{s}}}}]{Netopil_2022}
{Netopil}, M., {Oralhan}, {\.I}.~A., {{\c{C}}akmak}, H., {Michel}, R., \&
  {Karata{\c{s}}}, Y. 2022, \bibinfo{title}{{The Galactic metallicity gradient
  shown by open clusters in the light of radial migration},} \mnras, 509, 421,
  \dodoi{10.1093/mnras/stab2961}

\bibitem[{M. {Netopil} {et~al.}(2016){Netopil}, {Paunzen}, {Heiter}, \&
  {Soubiran}}]{Netopil_2016}
{Netopil}, M., {Paunzen}, E., {Heiter}, U., \& {Soubiran}, C. 2016,
  \bibinfo{title}{{On the metallicity of open clusters. III. Homogenised
  sample},} \aap, 585, A150, \dodoi{10.1051/0004-6361/201526370}

\bibitem[{C.~O. {Obasi} {et~al.}(2025){Obasi}, {Garro},
  {Fern{\'a}ndez-Trincado}, {Minniti}, {G{\'o}mez}, {Parisi}, \&
  {Ortigoza-Urdaneta}}]{Obasi_2025}
{Obasi}, C.~O., {Garro}, E.~R., {Fern{\'a}ndez-Trincado}, J.~G., {et~al.} 2025,
  \bibinfo{title}{{Multi-wavelength characterization of VVVX open clusters},}
  \aap, 695, A235, \dodoi{10.1051/0004-6361/202453227}

\bibitem[{K. {Pe{\~n}a Ram{\'i}rez} {et~al.}(2021){Pe{\~n}a Ram{\'i}rez},
  {Gonz{\'a}lez-Fern{\'a}ndez}, {Chen{\'e}}, \& {Ram{\'i}rez
  Alegr{\'i}a}}]{Pena_Ramirez_2021}
{Pe{\~n}a Ram{\'i}rez}, K., {Gonz{\'a}lez-Fern{\'a}ndez}, C., {Chen{\'e}},
  A.-N., \& {Ram{\'i}rez Alegr{\'i}a}, S. 2021, \bibinfo{title}{{The VVV open
  cluster project. Near-infrared sequences of NGC 6067, NGC 6259, NGC 4815,
  Pismis 18, Trumpler 23, and Trumpler 20},} \mnras, 503, 1864,
  \dodoi{10.1093/mnras/stab328}

\bibitem[{M.~S. {Pera} {et~al.}(2021){Pera}, {Perren}, {Moitinho}, {Navone}, \&
  {Vazquez}}]{Pera2021}
{Pera}, M.~S., {Perren}, G.~I., {Moitinho}, A., {Navone}, H.~D., \& {Vazquez},
  R.~A. 2021, \bibinfo{title}{{pyUPMASK: an improved unsupervised clustering
  algorithm},} \aap, 650, A109, \dodoi{10.1051/0004-6361/202040252}

\bibitem[{R.~L. {Phelps} {et~al.}(1994){Phelps}, {Janes}, \&
  {Montgomery}}]{Phelps_1994}
{Phelps}, R.~L., {Janes}, K.~A., \& {Montgomery}, K.~A. 1994,
  \bibinfo{title}{{Development of the Galactic Disk: A Search for the Oldest
  Open Clusters},} \aj, 107, 1079, \dodoi{10.1086/116920}

\bibitem[{A.~E. {Piatti} {et~al.}(1998){Piatti}, {Clari{\'a}}, {Bica},
  {Geisler}, \& {Minniti}}]{Piatti_1998}
{Piatti}, A.~E., {Clari{\'a}}, J.~J., {Bica}, E., {Geisler}, D., \& {Minniti},
  D. 1998, \bibinfo{title}{{A Photometric and Spectroscopic Study of the
  Southern Open Clusters PISMIS 18, PISMIS 19, NGC 6005, and NGC 6253},} \aj,
  116, 801, \dodoi{10.1086/300443}

\bibitem[{P. {Pi{\c{s}}mi{\c{s}}}(1959){Pi{\c{s}}mi{\c{s}}}}]{Pismis1959}
{Pi{\c{s}}mi{\c{s}}}, P. 1959, \bibinfo{title}{{Nuevos C{\'u}mulos Estelares en
  Regiones del Sur},} Boletin de los Observatorios de Tonantzintla y Tacubaya,
  2, 37

\bibitem[{O. {Plevne} \& F. {Akbaba}(2026){Plevne} \& {Akbaba}}]{Plevne2026}
{Plevne}, O., \& {Akbaba}, F. 2026, \bibinfo{title}{{Astrophysical Parameters
  of 5056 Open Star Clusters from Bayesian Nested Sampling with PARSEC
  Isochrones},} arXiv e-prints, arXiv:2605.23802,
  \dodoi{10.48550/arXiv.2605.23802}

\bibitem[{E. {Poggio} {et~al.}(2021){Poggio}, {Drimmel}, {Cantat-Gaudin},
  {Ramos}, {Ripepi}, {Zari}, {Andrae}, {Blomme}, {Chemin}, {Clementini},
  {Figueras}, {Fouesneau}, {Fr{\'e}mat}, {Lobel}, {Marshall}, {Muraveva}, \&
  {Romero-G{\'o}mez}}]{Poggio_2021}
{Poggio}, E., {Drimmel}, R., {Cantat-Gaudin}, T., {et~al.} 2021,
  \bibinfo{title}{{Galactic spiral structure revealed by Gaia EDR3},} \aap,
  651, A104, \dodoi{10.1051/0004-6361/202140687}

\bibitem[{S.~F. {Portegies Zwart} {et~al.}(2010){Portegies Zwart}, {McMillan},
  \& {Gieles}}]{Portegies-Zwart10}
{Portegies Zwart}, S.~F., {McMillan}, S. L.~W., \& {Gieles}, M. 2010,
  \bibinfo{title}{{Young Massive Star Clusters},} \araa, 48, 431,
  \dodoi{10.1146/annurev-astro-081309-130834}

\bibitem[{M.~F. {Qin} {et~al.}(2023){Qin}, {Zhang}, {Liu}, {Song}, {Hu},
  {Wang}, {Ma}, \& {L{\"u}}}]{Qin_2023}
{Qin}, M.~F., {Zhang}, Y., {Liu}, J., {et~al.} 2023, \bibinfo{title}{{Cluster
  aggregates surrounding Pismis 5 in the Vela molecular ridge},} \aap, 675,
  A67, \dodoi{10.1051/0004-6361/202244737}

\bibitem[{S. {Randich} {et~al.}(2022){Randich}, {Gilmore}, {Magrini}, {Sacco},
  {Jackson}, {Jeffries}, {Worley}, {Hourihane}, {Gonneau}, {Viscasillas
  Vazquez}, {Franciosini}, {Lewis}, {Alfaro}, {Allende Prieto}, {Bensby},
  {Blomme}, {Bragaglia}, {Flaccomio}, {Fran{\c{c}}ois}, {Irwin}, {Koposov},
  {Korn}, {Lanzafame}, {Pancino}, {Recio-Blanco}, {Smiljanic}, {Van Eck},
  {Zwitter}, {Asplund}, {Bonifacio}, {Feltzing}, {Binney}, {Drew}, {Ferguson},
  {Micela}, {Negueruela}, {Prusti}, {Rix}, {Vallenari}, {Bayo}, {Bergemann},
  {Biazzo}, {Carraro}, {Casey}, {Damiani}, {Frasca}, {Heiter}, {Hill},
  {Jofr{\'e}}, {de Laverny}, {Lind}, {Marconi}, {Martayan}, {Masseron},
  {Monaco}, {Morbidelli}, {Prisinzano}, {Sbordone}, {Sousa}, {Zaggia},
  {Adibekyan}, {Bonito}, {Caffau}, {Daflon}, {Feuillet}, {Gebran}, {Gonzalez
  Hernandez}, {Guiglion}, {Herrero}, {Lobel}, {Maiz Apellaniz}, {Merle},
  {Mikolaitis}, {Montes}, {Morel}, {Soubiran}, {Spina}, {Tabernero},
  {Tautvai{\v{s}}iene}, {Traven}, {Valentini}, {Van der Swaelmen}, {Villanova},
  {Wright}, {Abbas}, {Aguirre B{\o}rsen-Koch}, {Alves}, {Balaguer-Nunez},
  {Barklem}, {Barrado}, {Berlanas}, {Binks}, {Bressan}, {Capuzzo-Dolcetta},
  {Casagrande}, {Casamiquela}, {Collins}, {D'Orazi}, {Dantas}, {Debattista},
  {Delgado-Mena}, {Di Marcantonio}, {Drazdauskas}, {Evans}, {Famaey},
  {Franchini}, {Fr{\'e}mat}, {Friel}, {Fu}, {Geisler}, {Gerhard}, {Gonzalez
  Solares}, {Grebel}, {Gutierrez Albarran}, {Hatzidimitriou}, {Held},
  {Jim{\'e}nez-Esteban}, {J{\"o}nsson}, {Jordi}, {Khachaturyants},
  {Kordopatis}, {Kos}, {Lagarde}, {Mahy}, {Mapelli}, {Marfil}, {Martell},
  {Messina}, {Miglio}, {Minchev}, {Moitinho}, {Montalban}, {Monteiro},
  {Morossi}, {Mowlavi}, {Mucciarelli}, {Murphy}, {Nardetto}, {Ortolani},
  {Paletou}, {Palou{\v{s}}}, {Paunzen}, {Pickering}, {Quirrenbach}, {Re
  Fiorentin}, {Read}, {Romano}, {Ryde}, {Sanna}, {Santos}, {Seabroke},
  {Spagna}, {Steinmetz}, {Stonkut{\'e}}, {Sutorius}, {Th{\'e}venin}, {Tosi},
  {Tsantaki}, {Vink}, {Wright}, {Wyse}, {Zoccali}, {Zorec}, {Zucker}, \&
  {Walton}}]{Randich_2022}
{Randich}, S., {Gilmore}, G., {Magrini}, L., {et~al.} 2022,
  \bibinfo{title}{{The Gaia-ESO Public Spectroscopic Survey: Implementation,
  data products, open cluster survey, science, and legacy},} \aap, 666, A121,
  \dodoi{10.1051/0004-6361/202243141}

\bibitem[{M. {Riello} {et~al.}(2021){Riello}, {De Angeli}, {Evans},
  {Montegriffo}, {Carrasco}, {Busso}, {Palaversa}, {Burgess}, {Diener},
  {Davidson}, {Rowell}, {Fabricius}, {Jordi}, {Bellazzini}, {Pancino},
  {Harrison}, {Cacciari}, {van Leeuwen}, {Hambly}, {Hodgkin}, {Osborne},
  {Altavilla}, {Barstow}, {Brown}, {Castellani}, {Cowell}, {De Luise},
  {Gilmore}, {Giuffrida}, {Hidalgo}, {Holland}, {Marinoni}, {Pagani},
  {Piersimoni}, {Pulone}, {Ragaini}, {Rainer}, {Richards}, {Sanna}, {Walton},
  {Weiler}, \& {Yoldas}}]{Riello_2021}
{Riello}, M., {De Angeli}, F., {Evans}, D.~W., {et~al.} 2021,
  \bibinfo{title}{{Gaia Early Data Release 3. Photometric content and
  validation},} \aap, 649, A3, \dodoi{10.1051/0004-6361/202039587}

\bibitem[{J. Ruprecht(1966)Ruprecht}]{Ruprecht_1966}
Ruprecht, J. 1966, \bibinfo{title}{Classification of open star clusters,}
  Bulletin of the Astronomical Institutes of Czechoslovakia, 17, 33

\bibitem[{L. Sampedro {et~al.}(2017)Sampedro, Dias, Alfaro, Monteiro, \&
  Molino}]{Sampedro_2017}
Sampedro, L., Dias, W.~S., Alfaro, E.~J., Monteiro, H., \& Molino, A. 2017,
  \bibinfo{title}{{A multimembership catalogue for 1876 open clusters using
  {UCAC}4 data},} Monthly Notices of the Royal Astronomical Society, 470, 3937,
  \dodoi{10.1093/mnras/stx1485}

\bibitem[{D.~J. {Schlegel} {et~al.}(1998){Schlegel}, {Finkbeiner}, \&
  {Davis}}]{Schlegel1998}
{Schlegel}, D.~J., {Finkbeiner}, D.~P., \& {Davis}, M. 1998,
  \bibinfo{title}{{Maps of Dust Infrared Emission for Use in Estimation of
  Reddening and Cosmic Microwave Background Radiation Foregrounds},} \apj, 500,
  525, \dodoi{10.1086/305772}

\bibitem[{J. {Skilling}(2004){Skilling}}]{Skilling_2004}
{Skilling}, J. 2004, \bibinfo{title}{{Nested Sampling},} in American Institute
  of Physics Conference Series, Vol. 735, Bayesian Inference and Maximum
  Entropy Methods in Science and Engineering: 24th International Workshop on
  Bayesian Inference and Maximum Entropy Methods in Science and Engineering,
  ed. R.~{Fischer}, R.~{Preuss}, \& U.~V. {Toussaint} (AIP), 395--405,
  \dodoi{10.1063/1.1835238}

\bibitem[{J.~S. {Speagle}(2020){Speagle}}]{Speagle2020}
{Speagle}, J.~S. 2020, \bibinfo{title}{{dynesty: a dynamic nested sampling
  package for estimating Bayesian posteriors and evidences},} \mnras, 493,
  3132, \dodoi{10.1093/mnras/staa278}

\bibitem[{L. {Spina} {et~al.}(2022){Spina}, {Magrini}, \& {Cunha}}]{Spina22}
{Spina}, L., {Magrini}, L., \& {Cunha}, K. 2022, \bibinfo{title}{{Mapping the
  Galactic Metallicity Gradient with Open Clusters: The State-of-the-Art and
  Future Challenges},} Universe, 8, 87, \dodoi{10.3390/universe8020087}

\bibitem[{A.~L. {Tadross}(2008){Tadross}}]{Tadross_2008}
{Tadross}, A.~L. 2008, \bibinfo{title}{{A Reinvestigation of the Physical
  Properties of Pismis 3 based on 2MASS Photometry},} Chinese Journal of
  Astronomy and Astrophysics, 8, 362, \dodoi{10.1088/1009-9271/8/3/13}

\bibitem[{B. {Tan{\i}k {\"O}zt{\"u}rk} {et~al.}(2025){Tan{\i}k {\"O}zt{\"u}rk},
  {Bilir}, {Yontan}, {Plevne}, {Ak}, {Ak}, {Canbay}, \& {Banks}}]{Tanik2025}
{Tan{\i}k {\"O}zt{\"u}rk}, B., {Bilir}, S., {Yontan}, T., {et~al.} 2025,
  \bibinfo{title}{{A Comprehensive Study of Czernik 41 and NGC 1342 Using CCD
  UBV and Gaia DR3 Data},} \aj, 170, 164, \dodoi{10.3847/1538-3881/adefe0}

\bibitem[{R.~J. {Trumpler}(1930){Trumpler}}]{Trumpler_1930}
{Trumpler}, R.~J. 1930, \bibinfo{title}{{Preliminary results on the distances,
  dimensions and space distribution of open star clusters},} Lick Observatory
  Bulletin, 420, 154, \dodoi{10.5479/ADS/bib/1930LicOB.14.154T}

\bibitem[{N. {Vogt} \& A.~F.~J. {Moffat}(1973){Vogt} \& {Moffat}}]{Vogt_1973}
{Vogt}, N., \& {Moffat}, A.~F.~J. 1973, \bibinfo{title}{{Southern open clusters
  - II - UBV H-beta photometry of 11 clusters between galactic longitudes 259 -
  280.},} \aaps, 9, 97

\bibitem[{S. {Wang} \& X. {Chen}(2019){Wang} \& {Chen}}]{Wang2019}
{Wang}, S., \& {Chen}, X. 2019, \bibinfo{title}{{The Optical to Mid-infrared
  Extinction Law Based on the APOGEE, Gaia DR2, Pan-STARRS1, SDSS, APASS,
  2MASS, and WISE Surveys},} \apj, 877, 116, \dodoi{10.3847/1538-4357/ab1c61}

\bibitem[{T. {Yontan} {et~al.}(2026){Yontan}, {Bilir}, {\c{C}akmak}, {Plevne},
  {et~al.}}]{Yontan2026}
{Yontan}, T., {Bilir}, S., {\c{C}akmak}, H., {Plevne}, O., {et~al.} 2026,
  \bibinfo{title}{{Kings of the Milky Way: A Homogeneous Gaia DR3 Analysis of
  King Open Clusters and the Galactic Disc Metallicity Gradient},} arXiv
  e-prints

\bibitem[{S.~B. {Zhang} {et~al.}(2025){Zhang}, {Wei}, {Yang}, {Dai}, {Wang},
  {Toomey}, {Wang}, {Hobbs}, {Wu}, \& {Staveley-Smith}}]{Zhang_2025}
{Zhang}, S.~B., {Wei}, J.~J., {Yang}, X., {et~al.} 2025,
  \bibinfo{title}{{Searching for Radio Pulsars in Old Open Clusters from the
  Parkes Archive},} \apj, 988, 21, \dodoi{10.3847/1538-4357/ade802}

\end{thebibliography}

\end{document}